\documentclass[conference]{IEEEtran}
\IEEEoverridecommandlockouts
\usepackage{amsmath,amsfonts}
\usepackage{algorithm,algorithmic}
\usepackage{graphicx}
\usepackage{textcomp}
\usepackage{xcolor}
\usepackage{subfigure}
\usepackage{xspace}
\usepackage{tikz}
\usepackage{xcolor}
\usepackage{makecell}
\usepackage{bm}
\usepackage{balance}
\usepackage{hyperref} 
\usepackage[switch]{lineno} % <--- Add this line with the switch option
\usepackage[symbol]{footmisc}
\usepackage{multirow}

\newcommand{\abrar}[1]{{\color{black}{#1}}}

\newcommand{\clomp}{\texttt{Clomp}\xspace}
\newcommand{\kripke}{\texttt{Kripke}\xspace}
\newcommand{\lulesh}{\texttt{Lulesh}\xspace}
\newcommand{\hypre}{\texttt{Hypre}\xspace}

\newcommand{\ouralg}{$\mathsf{LASP}$\xspace}

\def\BibTeX{{\rm B\kern-.05em{\sc i\kern-.025em b}\kern-.08em
T\kern-.1667em\lower.7ex\hbox{E}\kern-.125emX}}

\begin{document}

\title{HPC Application Parameter Autotuning on Edge Devices: A Bandit Learning Approach\\
% {\footnotesize \textsuperscript{*}Note: Sub-titles are not captured in Xplore and should not be used}
}

\author{\IEEEauthorblockN{Abrar Hossain\textsuperscript{1}, Abdel-Hameed A. Badawy\textsuperscript{2}, Mohammad A. Islam\textsuperscript{3}, Tapasya Patki\textsuperscript{4}, Kishwar Ahmed\textsuperscript{1}}
\IEEEauthorblockA{\textit{\textsuperscript{1}Department of Electrical Engineering and Computer Science, The University of Toledo}\\
\textsuperscript{2}\textit{Department of Electrical and Computer Engineering, New Mexico State University}\\
\textsuperscript{3}\textit{Department of Computer Science and Engineering, The University of Texas at Arlington}\\
\textsuperscript{4}\textit{Center for Applied Scientific Computing, Lawrence Livermore National Laboratory}\\
Email: \textsuperscript{1}abrar.hossain@utoledo.edu,
\textsuperscript{2}badawy@nmsu.edu,
\textsuperscript{3}mislam@uta.edu,
\textsuperscript{4}patki1@llnl.gov,
\textsuperscript{1}kishwar.ahmed@utoledo.edu}
% \and
% \IEEEauthorblockN{Abdel-Hameed A. Badawy}
% \IEEEauthorblockA{\textit{New Mexico State University} \\
% badawy@nmsu.edu}
% \and
% \IEEEauthorblockN{Mohammad A. Islam}
% \IEEEauthorblockA{\textit{The University of Texas at Arlington} \\
% mislam@uta.edu}
% \and
% \IEEEauthorblockN{Tapasya Patki}
% \IEEEauthorblockA{\textit{Lawrence Livermore National Laboratory} \\
% patki1@llnl.gov}
% \and
% \IEEEauthorblockN{Kishwar Ahmed}
% \IEEEauthorblockA{\textit{The University of Toledo}\\
% kishwar.ahmed@utoledo.edu}
}

% \author{\IEEEauthorblockN{Abrar Hossain}
% \IEEEauthorblockA{\textit{The University of Toledo}\\
% abrar.hossain@rockets.utoledo.edu}
% \and
% \IEEEauthorblockN{Abdel-Hameed A. Badawy}
% \IEEEauthorblockA{\textit{New Mexico State University} \\
% badawy@nmsu.edu}
% \and
% \IEEEauthorblockN{Mohammad A. Islam}
% \IEEEauthorblockA{\textit{The University of Texas at Arlington} \\
% mislam@uta.edu}
% \and
% \IEEEauthorblockN{Tapasya Patki}
% \IEEEauthorblockA{\textit{Lawrence Livermore National Laboratory} \\
% patki1@llnl.gov}
% \and
% \IEEEauthorblockN{Kishwar Ahmed}
% \IEEEauthorblockA{\textit{The University of Toledo}\\
% kishwar.ahmed@utoledo.edu}
% }

\maketitle

\begin{abstract}
The growing necessity for enhanced processing capabilities in edge devices with limited resources has led us to develop effective methods for improving high-performance computing (HPC) applications. In this paper, we introduce \ouralg (\textbf{L}ightweight \textbf{A}utotuning of \textbf{S}cientific Application \textbf{P}arameters), a novel strategy designed to address the parameter search space challenge in edge devices. Our strategy employs a multi-armed bandit (MAB) technique focused on online exploration and exploitation. Notably, \ouralg takes a dynamic approach, adapting seamlessly to changing environments. We tested \ouralg with four HPC applications: \lulesh, \kripke, \clomp, and \hypre. Its lightweight nature makes it particularly well-suited for resource-constrained edge devices. By employing the MAB framework to efficiently navigate the search space, we achieved significant performance improvements while
adhering to the stringent computational limits of edge devices. Our experimental results demonstrate the effectiveness of \ouralg in optimizing parameter search on edge devices.

\end{abstract}


\section{Introduction}
\label{sec:intro}

\textbf{Motivation.} 
Edge devices have been gaining popularity as a platform to execute computational workloads for widespread availability and increasing computational power~\cite{shi2016edge}. According to a recent report~\cite{jaan2022pcie}, the market of edge-to-process application data is expected to grow by 75\% by 2026. Edge computing processes workload generated by end users nearby, thereby achieving low end-to-end latency and high bandwidth. 
High-performance computing (HPC) applications are characterized by their need for extensive computational resources and efficient performance. Edge devices can be used for scientific application execution due to their increasing processing capabilities. Recent U.S. DOE and Europe HPC reports~\cite{ beckman20205g} outline the opportunities to solve scientific applications on the backdrop of emerging edge computing technologies. However, limited and heterogeneous distributed edge resources present unique challenges to HPC execution on edge devices.

% \subsection{The problem}
\begin{figure}
    \centering
    \includegraphics[width=1\linewidth]{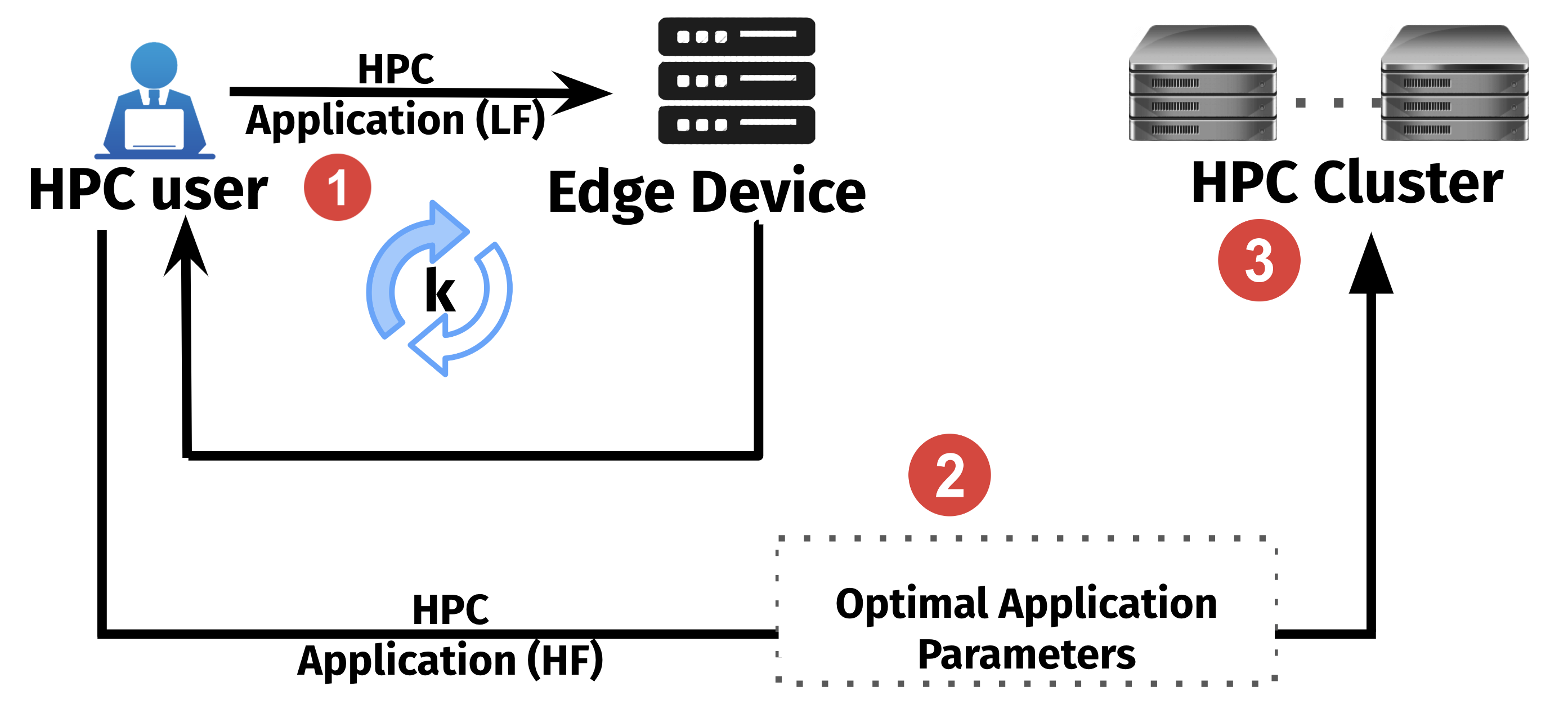}
    \vspace{-3mm}
    \caption{Framework to leverage edge devices to find the optimal parameters to execute applications on HPC clusters.}
    \label{fig:lasp_principle}
\end{figure}

HPC applications involve complex parameter configurations~\cite{tiwari2011auto}, which significantly affect their performance, contributing towards performance degradation and sometimes even causing non-execution faults~\cite{hu2020automated}. As such, it becomes challenging for the users to evaluate the impact of various tunable parameters on the execution time and understand their effects on each other~\cite{sarkar2009software}. Application users must invest considerable effort in searching for the optimal values for all parameters to attain the least execution time~\cite{silvano2016antarex}. Because manual tuning is time-consuming and labor-intensive and prone to significant error, the automatic tuning of configuration parameters for HPC applications has been a significant subject of study for the past several years~\cite{zhu2017bestconfig, chen2019d}.  \abrar{We propose an innovative approach where HPC applications are initially executed on edge devices to determine optimal application-level parameters. The edge devices can efficiently identify the best parameters by running these applications at low fidelity (LF), which demands fewer computational resources. These parameters are then transferred to traditional HPC platforms for execution at high fidelity (HF). This method significantly reduces the time and energy typically spent on parameter tuning on traditional HPC systems, leading to more efficient overall execution of HPC applications. Our approach is illustrated in Fig.~\ref{fig:lasp_principle}, where edge devices act as a preliminary stage for parameter optimization before the final execution on HPC clusters. 

Notably, existing parameter autotuning techniques have been developed primarily for traditional HPC systems, which themselves demand significant computational resources. Our motivating experiments on four HPC applications on edge devices show the unique challenges HPC parameter autotuning presents on edge platforms. By leveraging edge devices, this paper aims to enhance the efficiency and performance of traditional HPC applications. Our method, based on stochastic techniques for application-level parameters, is portable across various edge and HPC platforms, though some tuning may be required for hardware-level parameters.

% Addressing the HPC application parameter search challenges for edge devices is of paramount importance as there have been increasing interests in DOE projects~\cite{DBLP:conf/ieeesensors/BeckmanSCFJP16} to integrate HPC with edge computing.

} 
% We show the effectiveness of our proposed approach through experiments later in the paper.

% \subsection{Existing Solutions}

\textbf{Limitations of state-of-the-art approaches.}
Traditional parameter tuning methods are either exhaustive, time-consuming, or based on heuristics that may not capture the nuances of different application scenarios and the resource-constrained nature and volatility of the edge devices. Existing knowledge-based tuning involves domain experts manually adjusting parameters based on experience and intuition. While this can be effective, it is time-consuming, not scalable, and heavily relies on expert availability, while heuristic approaches utilize rule-based methods~\cite{kirkpatrick1983optimization, kennedy1995particle} to select parameters. These methods are faster but often need more flexibility to adapt to different application needs or changes in the computing environment, and thus, usually get stuck at local optima. Both manual and heuristic methods do not scale well with the increasing complexity of HPC systems~\cite{tiwari2011auto}.
  
To tackle these challenges, state-of-the-art solutions have employed variants of learning-based approaches~\cite{chen2018tvm, bei2015rfhoc}. Recently, the effectiveness of configuration autotuning has been demonstrated by more advanced learning techniques such as utilizing machine learning (ML) techniques~\cite{cheng_efficient_2021, yu2018datasize}. 
However, these models also come with their own overhead costs, making them non-ideal for edge devices. While numerous HPC applications may undergo multiple executions, the input type or size can vary over time. The optimal configuration evolves with changes in input type, input size, or the integration of incremental algorithmic improvements into the application code base~\cite{roy2021bliss}. Consequently, the cumulative cost of autotuning increases over time, and autotuning efforts may demand substantial resources on large-scale systems, resulting in the dedication of millions of node hours for autotuning on expensive supercomputers~\cite{ balaprakash_autotuning_2018}. Simultaneously, the correlation with workload type and input dataset size in big data applications fluctuates, leading to the frequent initiation of time-consuming model retraining tasks~\cite{sid2019multitask}. 

Predictive models can provide quicker solutions but often require substantial training data and are usually limited by the accuracy of their underlying models. They also face challenges in generalizing across different HPC applications and may require retraining for different environments~\cite{thiagarajan_bootstrapping_2018}. More importantly, these models are generally static, often leading to suboptimal performance or excessive computational costs~\cite{wood2021artemis}. HPC workloads and environments are highly dynamic; therefore, a tuning method that can adapt in real-time to changing conditions is required. However, existing predictive methods do not directly incorporate such dynamic workload in their learning~\cite{dou2023turbo}. 

Many search-based methods~\cite{krishna2020conex, zhu2017bestconfig} achieve satisfactory configuration for many HPC applications. These methods consider the relationship between performance and configuration parameters as a black box technique and employ a specific exploration mechanism to search for the optimal configuration directly. One prominent technique is the Bayesian Optimization (BO).
BO-based techniques and their variations can identify a near-optimal configuration with only a limited number of iterations for various HPC applications~\cite{ dou2020hdconfigor}.
However, the BO-based techniques have several limitations -- (1) Bayesian optimization struggles with the intricate relationship in big data frameworks, requiring numerous iterations for an accurate model~\cite{xin2022locat}; (2) Vanilla BO prioritizes quick convergence, risking time-consuming sub-optimal configurations due to overlooking evaluation times~\cite{dou2023turbo}; and (3) HPC workload characteristics change over time, necessitating configuration re-tuning, while, Vanilla BO lacks historical knowledge utilization and starts afresh for each task~\cite{dou2023turbo}. 

\textbf{Key Insights and Contributions.}
To address the limitations of existing approaches, we propose a novel lightweight and online technique for determining the optimal HPC configuration on resource-constrained edge devices: \textbf{L}ightweight \textbf{A}utotuning of \textbf{S}cientific Application \textbf{P}arameters (\ouralg). We focus on the challenges of configuration selection in HPC for edge devices.

Our solution leverages the multi-arm bandit (MAB) technique, offering unique benefits for HPC applications. \textit{First,} the flexibility of MAB models allows effective application across various HPC scenarios, adapting to specific needs and constraints. \textit{Second,} to our knowledge, we are the first to apply this approach to autotuning on edge devices. We compare \ouralg's autotuning effectiveness with the default strategy, where applications run with their default settings, demonstrating \ouralg's lightweight nature and minimal overhead. \textit{Third,} MAB models are adaptable, making them suitable for dynamic environments where reward distributions may change over time. This is particularly suitable to the volatile edge environment we are leveraging. Our performance evaluation shows that \ouralg\ can identify the best configuration, significantly enhancing HPC application performance on edge devices. Furthermore, our model dynamically adapts to user needs and changes in application behavior, determining the optimal configuration with minimal regret, thus fulfilling MAB properties.

% A Multi-Armed Bandit is a compelling solution to address these issues. It efficiently balances exploring various configurations and exploiting the best-found settings, thereby potentially reducing the tuning time while optimizing performance. The configuration of these applications involves numerous application-level parameters, which can extend from compile-time parameters to runtime parameters. Finding the optimal configuration from this intricate parameter space is thus a difficult task.
% High-Performance Computing (HPC) has traditionally relied on various methods for parameter tuning.

\textbf{Organization of the paper.}
The rest of this paper is organized as follows.
In Section~\ref{sec:background}, we present the background and discuss the challenges.
% In Section~\ref{sec:related}, we discuss the related studies that are most pertinent to this work.
In Section~\ref{sec:prob_form}, we formulate the problem and present the objective function. 
% We formulate the problem and present an efficient algorithm.
In Section~\ref{sec:alg}, we introduce a lightweight technique for HPC application parameter selection. 
In Section~\ref{sec:evaluation}, we present results to show performance evaluation in dynamic workload scenarios.  
In Section~\ref{sec:conclude}, we conclude our paper and suggest future directions.

\vspace{-2mm}
\section{Preliminaries}
\label{sec:background}

% \subsection{Definitions}
\subsection{Terminology}
% This paper defines essential terms before quantifying the challenges associated with creating an effective auto-tuner.
We first define essential terms that are used throughout the paper.
\textbf{Tunable parameters} include the application-level parameters, which can take on various values or states, markedly affecting the execution time of an application.

The \textbf{autotuning search space}, or \textbf{search space}, comprises the extensive $n$-dimensional space created by the range of values that tunable parameters can take. The range of this search space is defined by the potential combinations of tunable parameter configurations, represented by the product of each parameter’s possible values $(a_1 × a_2 × ... × a_n$, where $n$ denotes the number of tunable parameters).

A \textbf{configuration}, or a sample, is a specific combination of parameter values selected within the search space. Sampling or sample evaluation involves running an application using a particular configuration and assessing its runtime. \textbf{Oracle configuration} describes the ideal configuration with minimal execution time or power consumption. While it is intuitive to aim for shorter execution times, we also consider parameter configurations that minimize power consumption of edge device. This is because power is often a limited resource for edge devices, and optimizing for power efficiency is crucial to ensure their effective operation. Identifying the Oracle configuration accurately involves examining all possible configurations in the search space, which is impractical in production settings. However, we conduct an exhaustive search to assess the effectiveness of any given configuration relative to the Oracle configuration. This assessment is quantified as the distance from the Oracle configuration and is defined as follows:
\[
\left( \frac{\text{execution time of a configuration}}{\text{execution time of the Oracle configuration}} - 1 \right) \times 100\%.
\]

\subsection{Multi-Arm Bandit}
% Multi-armed bandit-based approaches provide a compelling and unique solution to address the issues. Unlike static models, it is designed to excel in dynamic environments. 
The multi-arm bandit (MAB) problem~\cite{slivkins2019introduction} is fundamental in probability theory and decision-making under uncertainty. It involves a sequential decision-making framework where an agent must choose with limited information. 
Pure exploration bandit problems aim to minimize the simple regret, defined as the distance from the optimal solution, as quickly as possible in any given setting. 
The pure-exploration MAB problem has a long history in the stochastic setting~\cite{bubeck2011pure}, and was recently extended to the non-stochastic setting~\cite{jamieson2016non}. Similarly, the stochastic pure-exploration infinite-armed bandit problem was studied by Carpentier et al.~\cite{carpentier2015simple}, where a pull of each arm $i$ yields an i.i.d. sample in $[0, 1]$ with expectation $\nu_i$, and $\nu_i$ is a loss drawn from a distribution with cumulative distribution function $F$. 
Hyperband~\cite{li2017hyperband} works by the best arm identification, i.e., selection of an arm with the highest average payoff in a non-stochastic setting.

The MAB technique has been applied in solving many real-life problems, including exploration and identification of efficient setting from a given distribution.
Some application domains include healthcare, finance~\cite{shen2015portfolio}, recommender systems, etc. Naturally, due to their ability to continuously learn and adapt their strategies based on real-time feedback, these approaches have also seen widespread adoption in hyperparameter tuning solutions for neural Networks~\cite{snoek2014input}.

In its basic stochastic form, the bandit problem involves a set of \( K \) probability distributions, denoted as \( \{D_1, \ldots, D_K\} \), each with associated expected values \( \{\mu_1, \ldots, \mu_K\} \) and variances \( \{\sigma^2_1, \ldots, \sigma^2_K\} \). Initially, these distributions are unknown to the player. These distributions are often likened to the arms of a slot machine, with the agent acting as a gambler whose goal is to maximize rewards by pulling these arms over multiple turns. At each turn \( t = 1, 2, \ldots \), the player chooses an arm, indexed by \( j(t) \), and receives a reward \( r(t) \sim D_{j(t)} \). The player’s objective is to determine which distribution has the highest expected value and accumulate as much reward as possible. Bandit algorithms guide the player in choosing an arm \( j(t) \) at each turn. The primary metric for evaluating these algorithms is the total expected regret, defined for a given turn \( T \) as:
\[ R_T = T \mu^* - \sum_{t=1}^{T} \mu_{j(t)}, \]
where \( \mu^* = \max_{i=1,\ldots,K} \mu_i \) is the expected reward from the best arm. Alternatively, the total expected regret can also be expressed as:

\begin{equation}
R_T = T \mu^* - \mu_{j(t)}\sum_{k=1}^{K} E[T_k(T)] \label{eq:regret_eqn},
\end{equation}
where \( T_k(T) \) is a random variable denoting the number of times arm \( k \) is played during the first \( T \) turns.

% They continuously learn and adapt their strategies based on real-time feedback, ensuring more effective tuning across varying conditions. By balancing the exploration of new configurations with the exploitation of known suitable configurations, MAB algorithms can efficiently navigate the vast parameter space, avoiding the computational intensity of brute-force methods. 

% where the objective of MAB algorithms is often to minimize regret, which is the difference between the rewards actually obtained and the rewards that could have been obtained if the best arm had been chosen every time. 

\subsection{Edge Devices as a Surrogate for Autotuning}
The use of edge devices for running HPC applications is increasingly gaining attention. The Waggle sensor platform~\cite{DBLP:conf/ieeesensors/BeckmanSCFJP16} is a key example of integrating HPC with edge computing, offering real-time data analysis and modular sensor network capabilities. An extension of this project, The Sage Continuum~\cite{SageContinuum2013} offers a distributed, software-defined sensor network that leverages machine learning and edge computing and provides a robust framework for real-time data analysis and sensor management. Bhupendra A. Raut et al.~\cite{raut2023optimizing} provides critical insights into optimizing algorithms for edge-computing sensor systems, particularly focusing on the stability and performance of the blockwise Phase Correlation method in estimating cloud motion vectors. Kim et al.~\cite{kim2022goal} introduces a two-layered scheduling model for edge computing and incorporates “science goals” to align user objectives with resource allocation, thereby offering a nuanced approach to HPC applications in edge systems. The Interconnected Science Ecosystem (INTERSECT) architecture open architecture~\cite{engelmann22intersect} is a federated instrument-to-edge-to-center framework that advocates autonomous data handling and processing in scientific research. This architecture aligns closely with the objectives of running HPC applications in edge systems and offers a system of systems and microservice architecture for enhanced scalability and adaptability.  

\begin{figure}[t!]
\subfigure[]{\label{fig:background_distance_from_oracle}\includegraphics[width=0.23
\textwidth]{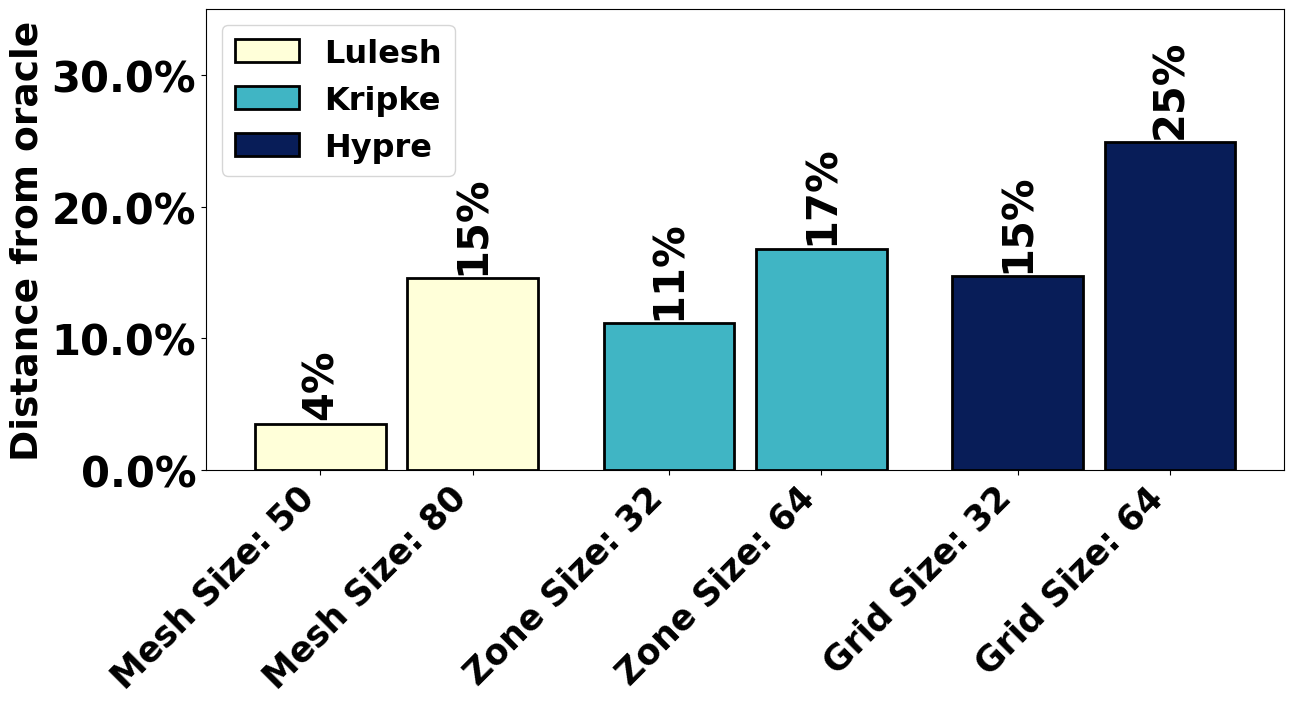}}\hspace{0.2cm}
\subfigure[]{\label{fig:num_of_common_configs}\includegraphics[width=0.23\textwidth]{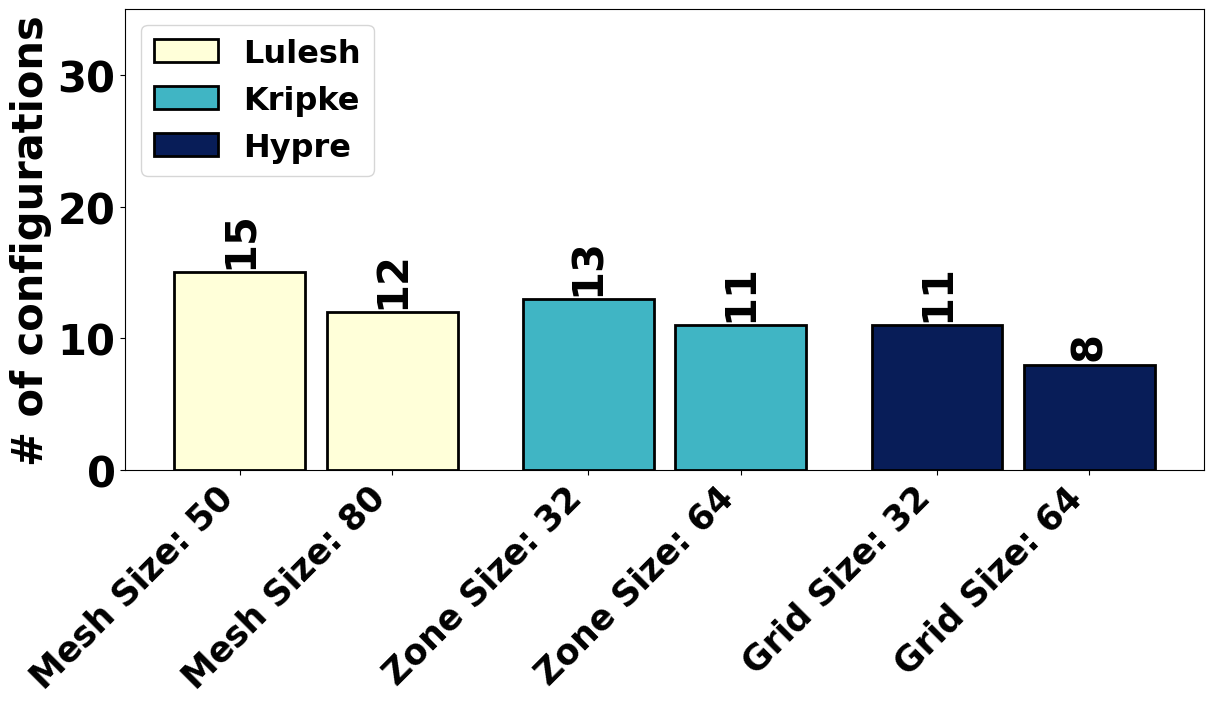}}
\vspace{-3mm}
\caption{Overlap of optimal configurations on low- and high-fidelity setting. (a) The top 20 configurations identified in the low-fidelity setting are compared to the optimal configuration when run on the high-fidelity setting of the target device, and the average distance between them is measured. (b) The number of common configurations out of top 20 configurations for both the low-fidelity and high-fidelity settings.}
\label{fig:number_of_configs_in_common}
\end{figure}

By processing data close to the source, edge computing can significantly reduce latency and bandwidth requirements, crucial for time-sensitive HPC applications\cite{hossain2010automating}. Edge devices also enable real-time data processing, which is essential for applications requiring immediate analysis and decision-making.  However, these unique advantages present their unique challenges as well. Unlike traditional supercomputing centers, edge devices suffer from limited computational power and memory, posing a challenge for resource-intensive HPC applications. The performance of edge devices can be inconsistent due to their varying specifications and the dynamic nature of edge environments.

\abrar {Our proposed HF/LF approach is designed to overcome this challenge and accommodate the dynamic nature of HPC workload and edge environment on which we are performing this autotuning task. Notably, our algorithm, \ouralg, is application-agnostic, meaning it can be employed with any application that associates distinct values with its parameters. In a multi-fidelity context, an application can be executed with varying levels of fidelity settings, such as adjusting the resolution in a numerical simulation or modifying the depth of a machine learning model. For example, the fidelity levels of \hypre is determined by the discretization using \( m^3 \) grid points, where \( m \) varies from \( m_{\text{min}} = 10 \) to \( m_{\text{max}} = 100 \). Due to the algebraic multigrid algorithm's computational complexity of \( O(m^3) \), the mapping from the fidelity parameter \( q \) to \( m \) is represented as a linear interpolation between \( [q_{\text{min}}, m^3_{\text{min}}] \) and \( [q_{\text{max}}, m^3_{\text{max}}] \). It is to be noted that, there is a trade-off in accuracy due to the shift between low and high fidelity levels, as lower fidelity runs on edge devices are inherently less accurate than those at higher fidelities on traditional HPC systems. However, this trade-off is acceptable, as we are not concerned with the specific results from the low-fidelity runs. Our primary goal is to use these low-fidelity edge device runs to effectively tune the parameters of the model. Importantly, our analysis in Fig.~\ref{fig:number_of_configs_in_common} shows that there is a significant overlap between the optimal parameters for both low and high fidelity settings, meaning that the parameters tuned at low fidelity are often effective at high fidelity as well.

We represent fidelity levels using \( q \in [q_{\text{min}}, q_{\text{max}}] \), where \( q_{\text{min}} \) and \( q_{\text{max}} \) indicate the minimum and maximum fidelity values, respectively. The time required for function evaluation is assumed to increase linearly with fidelity \( q \). To optimize efficiency and reduce tuning costs, we utilize lower fidelity settings on edge devices, leveraging their faster, lower-cost performance. These lower fidelity evaluations, \( g(y, q) \) where \( q < q_{\text{max}} \), serve as approximations of the high-fidelity objective function, \( g(y, q_{\text{max}}) \), which runs on traditional HPC systems. The overall goal is to determine the best tuning parameters \( y \) to optimize the high-fidelity function \( g(y, q_{\text{max}}) \) by using the lower fidelity, edge-based evaluations as proxies, thereby improving the efficiency of HPC tuning.

By leveraging this property, \ouralg can dynamically navigate the parameter space to identify the optimal configuration, regardless of the specific application. To address the dynamic environment, \ouralg incorporates a reward feedback mechanism, enabling the algorithm to operate in real-time and adapt to changing environments. We simulate this dynamic behavior by tuning four HPC applications, and introducing error measurements into our readings, as described in Section \ref{sec:evaluation}. Furthermore, in the same section, we demonstrate that our algorithm can yield satisfactory results under varying levels of power and CPU capping, underscoring its robustness and adaptability.}

%Our proposed MAB-based approach is also well-suited to address these challenges and make edge devices appropriate for HPC application execution.

In this study, we run applications on varying fidelity settings, for example, \lulesh (mesh size = 50, 80), \kripke (Zone size = 32, 64), and \hypre (Grid size  = 32, 64). In Fig.~\ref{fig:number_of_configs_in_common}(b), we see a significant overlap with the most optimal configurations compared to running them in a low and high-fidelity setting. As shown in Fig. \ref{fig:number_of_configs_in_common}(a), we observed that the top 20 configurations identified through low-fidelity simulations and then transferred to a high-fidelity setting achieved performance within 25\% of the optimal configuration (oracle) on the target device.

%Building on this, our Multi-Armed Bandit (MAB)-based framework utilizes a reward feedback mechanism to harness these low-fidelity results and continually adjust HPC application configurations. Through iterative interactions with the edge environment, the framework employs the MAB algorithm to efficiently exploit the advantages of edge computing, such as reduced latency, while mitigating its inherent limitations. By using the MAB approach to find optimal parameters cheaply on edge devices, we can make informed, intelligent decisions about resource allocation and workload distribution—key factors for optimizing the performance of HPC applications in edge environments.

% \end{flushleft}

\subsection{Challenges in HPC Parameter Search}
% \begin{flushleft}

% \subsection{Parameter Configuration}
% % \begin{flushleft}
% Tunable parameters encompass both application- and hardware-level variables, allowing for assigning multiple values or states. The execution time of an application is directly impacted by the specific values assigned to these tunable parameters.

% The optimization of parameter search space of an HPC application encompasses an \(n\)-dimensional space, where \(n\) represents the number of tunable parameters. The search space size corresponds to the total number of possible configurations resulting from combining the tunable parameters. With \(n\) parameters, which have \(a_i\) possible values for the \(i\)-th parameter, the size of the search space comes to a total of \(a_1 \times a_2 \times \ldots \times a_n\). A parameter configuration, referred to in this paper as a configuration or sample, represents a single permutation of all valid parameter values taken from the complete parameter search space.

% Now, we delve into the challenges of attaining an efficient parameter search optimization.
Numerous challenges are associated with attaining an efficient parameter search optimization.
\textit{First}, finding the optimal set of parameters necessitates exhaustive exploration within a vast multidimensional search space. For example, popular HPC applications, such as Kripke~\cite{kunen2015kripke} and AMG~\cite{richards2018quantitative}, have more than 1 million and 5 million tunable hardware and software parameters, respectively. Searching for the most optimal parameters from this vast set is infeasible without an efficient algorithm. \textit{Second}, conventional parameter search approaches often yield suboptimal configurations, highlighting the importance of capturing the interplay between application- and hardware-level tuning parameters to achieve maximum performance.

% \begin{figure}[ht]
%     \centering
%     \begin{subfigure}{0.23\textwidth}
%         \centering
%         \includegraphics[width=\linewidth]{figures/kripke_two_param_heatmap.png}
%         \caption{ Kripke runtime variability}
%         \label{fig:kripke_heatmap}
%     \end{subfigure}
%     \hfill
%     \begin{subfigure}{0.23\textwidth}
%         \centering
%         \includegraphics[width=\linewidth]{figures/diff_param_kripke_exec_time.png}
%         \caption{ Distance from oracle}
%         \label{fig:oracle_distance}
%     \end{subfigure}
%     \caption{The CDF of execution times of 500 randomly selected AMG configurations shows that the execution times are far from the Oracle. AMG’s execution times vary even when only two parameters are varied (others are fixed).}
%     \label{fig:two_figures}
% \end{figure}

\begin{figure}[t!]
\subfigure[]{\label{fig:kripke_two_param_heatmap}\includegraphics[width=0.23\textwidth]{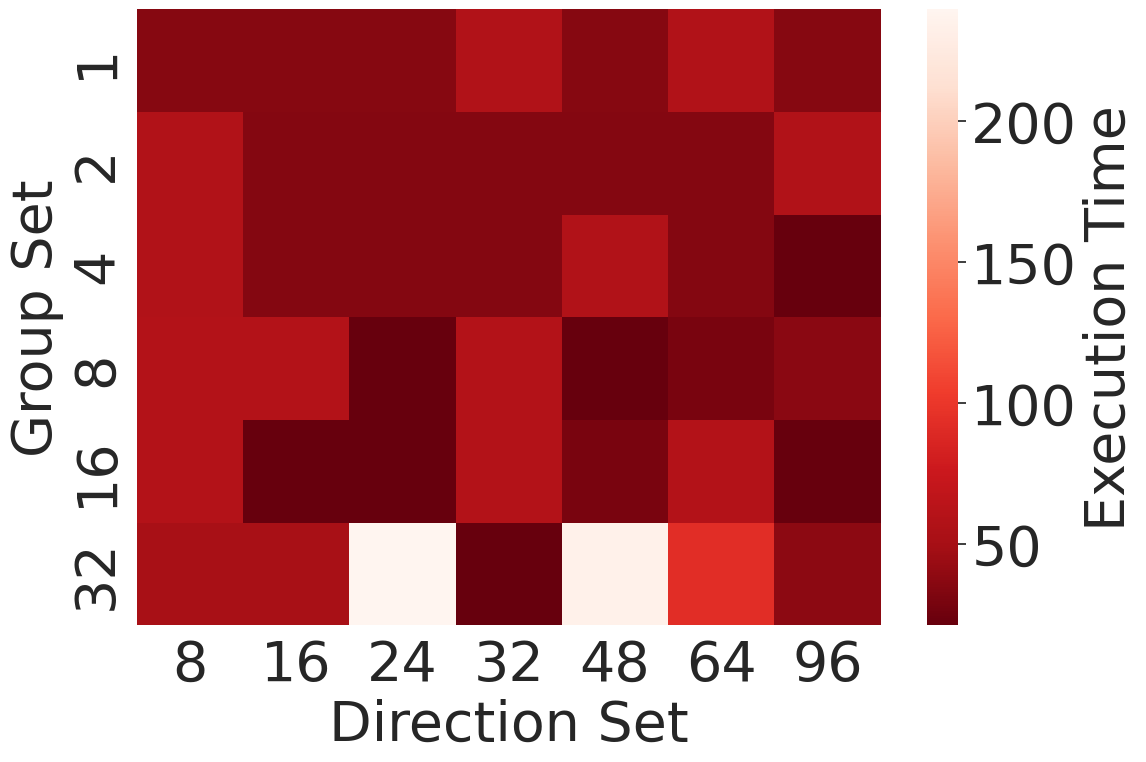}}\hspace{0.2cm}
\subfigure[]{\label{fig:varying_input_params_oracle}\includegraphics[width=0.23\textwidth]{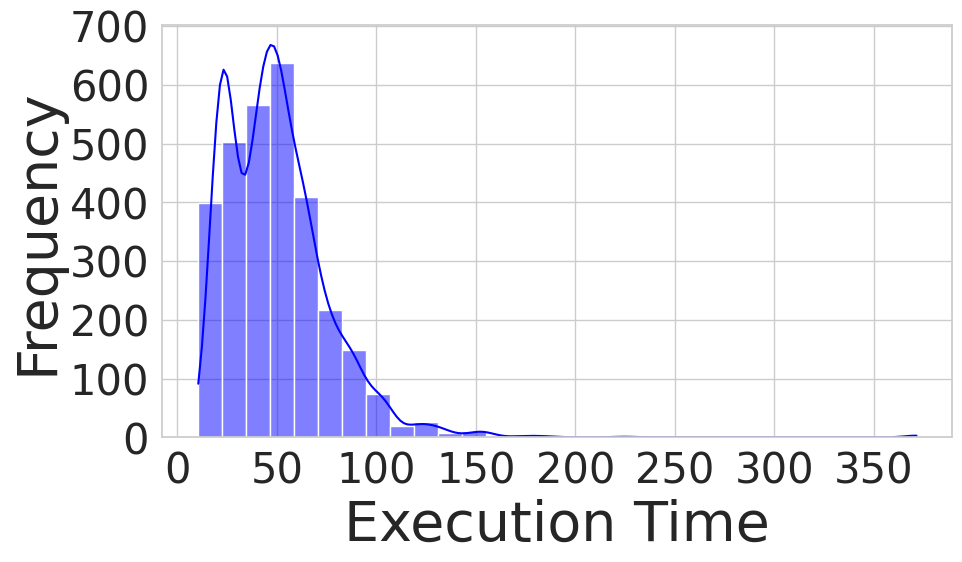}}
\vspace{-3mm}
\caption{Distribution of execution time for \kripke for all sets of configurations. (a) Tuning only two sets of parameters gives wide variance in the execution time. (b) Distribution of execution time for \kripke for all sets of configurations.}
\label{fig:kripke_execution}
\end{figure}
%\vspace{-4ex}
% \begin{figure}[t]
%   \centering
%   \begin{subfigure}{0.23\textwidth}
%     \includegraphics[width=\linewidth]{figures/kripke_two_param_heatmap.png}
%     \caption{Distribution of execution time for Kripke for all sets of configurations.}
%     \label{fig:kripke_two_param_heatmap}
%   \end{subfigure}
%   \begin{subfigure}{0.23\textwidth}
%     \includegraphics[width=\linewidth]{figures/varying_input_params_oracle.png}
%     \caption{Distribution of execution time for Kripke for all sets of configurations.}
%     \label{fig:varying_input_params_oracle}
%   \end{subfigure}%
%  \caption{}
%  \label{fig:kripke_execution}
% \end{figure} 

% \begin{figure}[H]
%   \centering
%   \includegraphics[width=\columnwidth]{figures/kripke_two_param_heatmap.png}
%   \caption{Performance Gain for different application}
%   \label{fig:varying_input_params_oracle}
% \end{figure}

% \begin{figure}[H]
%   \centering
%   \includegraphics[width=\columnwidth]{figures/varying_input_params_oracle.png}
%   \caption{Performance Gain for different application}
%   \label{fig:varying_input_params_oracle}
% \end{figure}

\begin{figure}
  \centering
\includegraphics[width=0.62\textwidth]{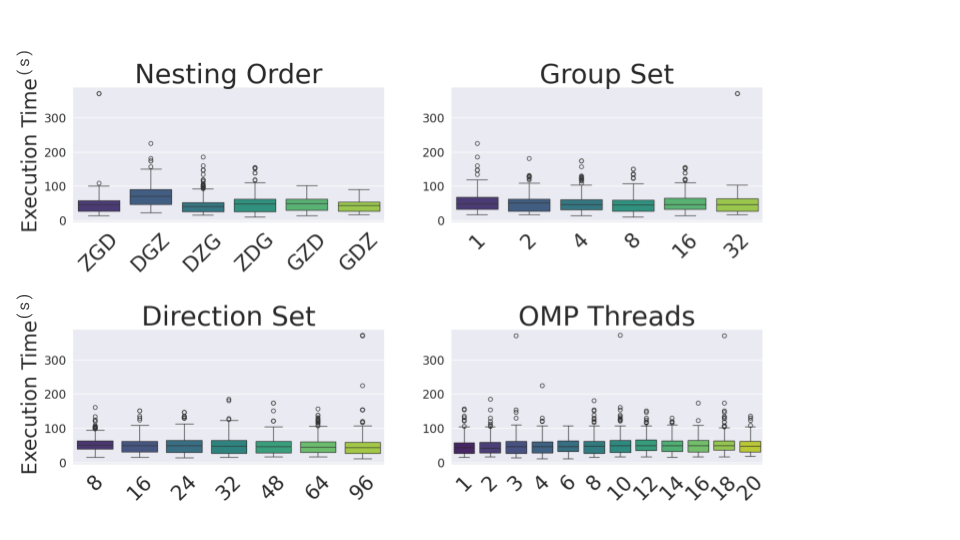}
  \vspace{-6mm}  
  \caption{Runtime variability of Kripke for different parameters considered independently.}
  \label{fig:4_param_kripke_exec}
\end{figure}

The significant impact of selecting the right configuration on the application’s execution time is demonstrated in Fig.~\ref{fig:kripke_two_param_heatmap}. This figure illustrates the variation in execution times that results from altering only two application-level parameters while keeping all other parameters constant. It is observed that the variance in execution time becomes much more pronounced when more parameters are modified. Fig.~\ref{fig:4_param_kripke_exec} illustrates the varying execution times resulting from tuning each parameter individually, reinforcing this key point. Additionally, Fig.~\ref{fig:varying_input_params_oracle} provides a distribution of execution times across all sets of configurations. 

This clearly highlights the crucial role of proper configuration selection in achieving optimal runtime performance.
Considering that most configurations deviate significantly from the absolute best-performing configuration, it is plausible to hypothesize that the challenge posed by a large search space can be mitigated by swiftly discarding the low-performing configurations, namely configurations with high runtimes. However, identifying these areas proves to be a formidable task, often fraught with the risk of overlooking the optimal configuration.

% \vspace{-3ex}
\section{Problem Formulation}
\label{sec:prob_form}

We assume an independently and identically distributed (i.i.d) rewards model, denoted as stochastic bandits. In our model, we assume a choice of \( K \) actions, which we refer to as arms, and which are to be executed over \( T \) rounds, where \( K \) and \( T \) are predefined. During every round, the algorithm selects one arm, leading to the accumulation of a reward specific to that arm. The primary objective of the algorithm is to optimize the total reward accumulated throughout the \( T \) rounds. The model includes the following assumptions. \emph{First}, we can only observe the reward associated with the action it chooses and no other information. Specifically, the model needs to be made aware of the potential rewards from other actions that were not selected (a.k.a., bandit feedback~\cite{slivkins2019introduction}). \emph{Second}, the reward corresponding to each action is i.i.d. For any given action ``a'', we assume a reward distribution, \( D_a \), over the real numbers. Each time we choose an action, its reward is independently drawn from \( D_a \). Initially, these reward distributions are unknown to the algorithm. \emph{Third}, we assume that the rewards received in each round are constrained within the range [0, 1].

%%%%%%
We include user-defined priorities when selecting the optimal HPC configuration. To include users in the decision framework, we include two parameters --  $\alpha$ for execution time and $\beta$ for power consumption, both ranging within [0, 1]. The user can set these parameters to control the optimization balance, e.g., higher values in $\alpha$ and $\beta$ indicate higher emphasis on execution time or power consumption, respectively.
% We design our Multi-Armed Bandit Algorithm for Optimal Configuration Selection, uniquely designed to optimize system configurations based on user-defined priorities in execution time and power consumption. This user-centric approach is facilitated by two parameters: $\alpha$ for execution time and $\beta$ for power consumption, both ranging within [0, 1]. 
%%%%
In our model, we define $\chi$ as the parameter space, where $\chi = \{1, \ldots, x\}$ is a finite action space; i.e., we set every unique combination of the parameters (configuration) as an arm of the MAB setting. 

% We formulate problem-based on the multi-arm bandit technique, where 
We specify a distribution $D$ over pairs $(x, ~r)$, where $x \in X$ denotes the parameter configuration and $r \in [0, 1]^A$ denotes a vector of rewards.  
In its basic stochastic form, our formulation involves a set of \( K \) probability distributions for each arm, denoted as \( \{D_1, \ldots, D_K\} \), each with associated expected values \( \{\mu_1, \ldots, \mu_K\} \) and variances \( \{\sigma^2_1, \ldots, \sigma^2_K\} \). Initially, these distributions are unknown to the algorithm. During each turn \( t = 1, 2, \ldots \), the algorithm chooses an arm, indexed by \( j(t) \), and receives a reward \( r(t) \sim D_{j(t)} \). The objective is to determine the distribution with the highest expected value and to accumulate as much reward as possible in each iteration. 

To model uncertainty, we employ an upper confidence bound (UCB)~\cite{auer2002using} technique that employs ``optimism under uncertainty''. Based on current observations, this technique assumes that every arm represents the best possible outcome. Consequently, the selection of an arm is based on these optimistic estimations. The technique involves initially trying each arm once. Then, for each round, \( t = 1, \ldots, T \), the technique selects the arm \( x(t) \) that appears to be the most promising. The selection of configurations in each iteration is calculated as follows for a configuration $x$ at iteration $t$:
\begin{align}
UCB(x, t) = R_x + \sqrt{\frac{2 \ln t}{N_x}},
\label{eqn:upper_confidence_bound}
\end{align}
where $R_x = f_{\text{reward}}(x)$ is the weighted reward for configuration $x$, and $N_x$ is the count of times configuration $x$ has been selected up to iteration $t$. Eq.~\ref{eqn:upper_confidence_bound} dynamically balances the exploration of new configurations against exploiting those already known to be effective. The proposed model ensures that the reward is inversely proportional to the normalized metrics of execution time and power consumption, thereby aligning with the user's optimization goals.
%%%%
% We evaluate each configuration in the space $\chi$ based on two key metrics: execution time ($\tau_x$) and power consumption ($\rho_x$). The algorithm operates over a set number of iterations, denoted as $T$. During each iteration $t$, it selects a configuration $x_t$ from $\chi$. A calculated reward guides this selection, a function of the user-specified parameters $\alpha$ and $\beta$, as well as the normalized execution time and power consumption values.

% The reward for selecting a configuration $x$ at iteration $t$, denoted as $R_{x,t}$, is computed using the formula:

% \[ f_{\text{reward}}(x) = \alpha \times \left( \frac{1}{\mu(\tau_{x})} \right) + \beta \times \left( \frac{1}{\mu(\rho_{x})} \right) \]

% The selection of configurations in each iteration is guided by a modified Upper Confidence Bound (UCB) formula. The UCB for a configuration $x$ at iteration $t$ is calculated as follows:

% \[ UCB(x, t) = R_x + \sqrt{\frac{2 \ln t}{N_x}} \]

% In this formula, $R_x = f_{\text{reward}}(x)$ is the weighted reward for configuration $x$, and $N_x$ is the count of times configuration $x$ has been selected up to iteration $t$. This UCB calculation dynamically balances the exploration of new configurations against exploiting those already known to be effective.

After each iteration $t$, we identify the configuration $x$ with the highest UCB value. The configuration, $x_t^*$, is determined as follows:
\begin{align}
x_t^* = \arg\max_x UCB(x, t).
\end{align}

This iterative selection strategy ensures an adaptive balance between exploring untested configurations and exploiting known effective ones. We determine the most frequently selected configuration as follows: 
\begin{align}
x_{\text{opt}} = \arg\max_x N_x.
\end{align}

\begin{figure}
    \centering
    \includegraphics [width=1\linewidth]{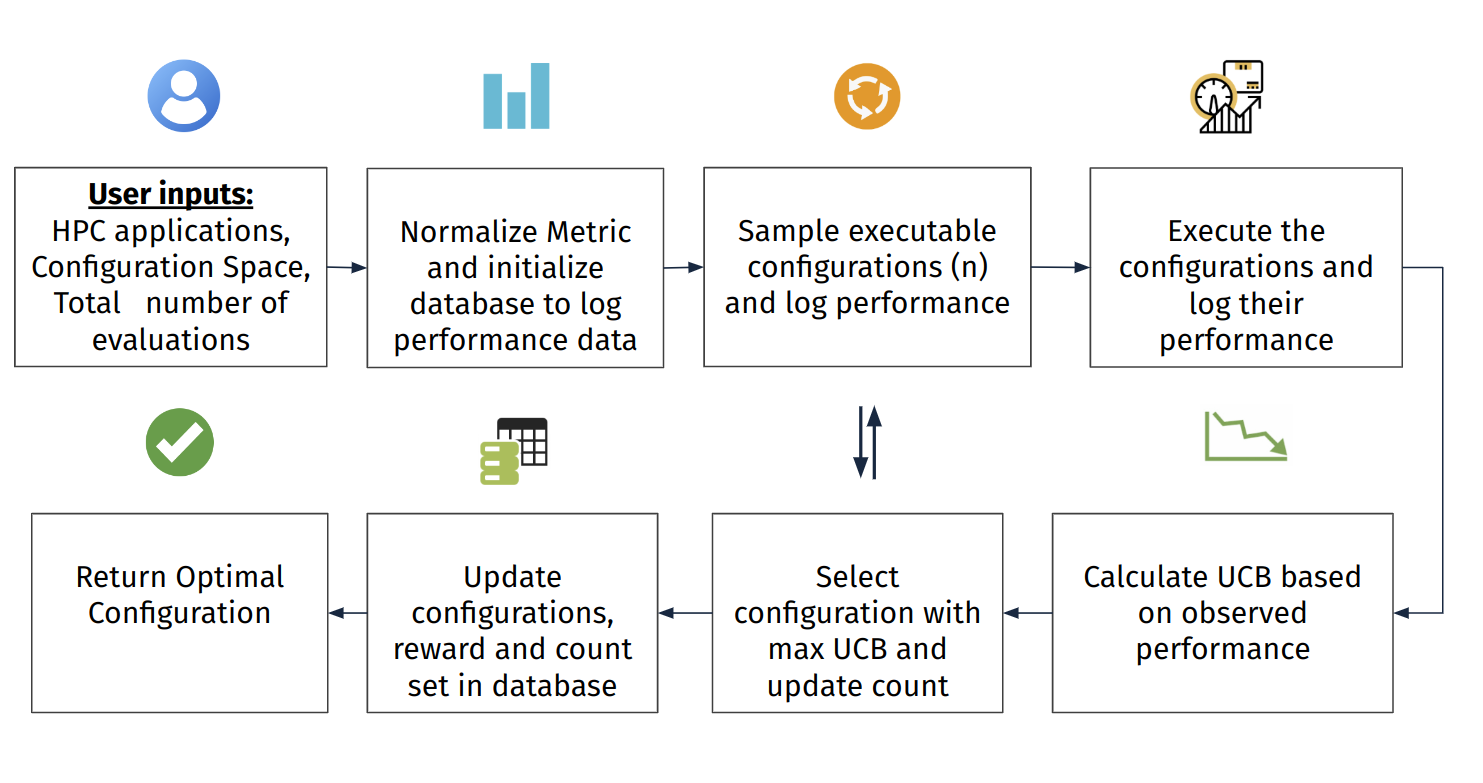}
    \vspace{-7mm}
    \caption{Block diagram of the \ouralg.}
    \label{fig:block_diagram}
\end{figure}

After $T$ round of iterations, the algorithm outlined in Section~\ref{sec:alg} outputs the most optimal configuration, $x_{\text{opt}}$. The high-level block diagram of \ouralg is given in Fig.~\ref{fig:block_diagram}.

% \kishwar{The high-level block diagram of \ouralg is given in Fig.~\ref{fig:block_diagram}.}

% The ultimate objective of our algorithm is to maximize this expected total reward, aligning with the user-defined optimization criteria set by $\alpha$ and $\beta$. By allowing the user to specify these parameters, the algorithm offers a tailored approach to configuration selection, efficiently addressing the unique requirements of each use case.
% \section{HPC Parameter Autotuning}

\section{Lightweight Autotuning of HPC Application Parameter}
\label{sec:alg}

\begin{algorithm}[t!]
  \caption{Lightweight Autotuning of Scientific Application Parameters (\ouralg)}
  \label{alg:mab_algorithm}
  \textbf{Input:} Configuration space ($\chi$), total iterations ($T$), execution time weight parameter ($\alpha$), and power consumption weight parameter ($\beta$)
  \begin{algorithmic}[1]
  \STATE \textbf{Initialization:} Dictionary for counting selections of each configuration ($N_x$), reward metrics ($\tau$ and $\rho$)
    \STATE Apply MinMax normalization: $\tau \leftarrow \frac{\tau - \min(\tau)}{\max(\tau) - \min(\tau)}, \rho \leftarrow \frac{\rho - \min(\rho)}{\max(\rho) - \min(\rho)}$
    \STATE \textbf{for} {$t \in \{1, 2, \ldots, T\}$} \textbf{do}
    \STATE \quad  \textbf{for}  configuration {$x \in \chi$} \textbf{do}
    \STATE \quad \quad Calculate weighted reward $R_x = w_{\tau} \times \left( \frac{1}{\mu(\tau_x)} \right) + w_{\rho} \times \left( \frac{1}{\mu(\rho_x)} \right)$

    \STATE \quad \quad Calculate UCB values for each configuration using:
    \STATE \quad\quad $UCB(x, t) = R_x + \sqrt{\frac{2 \ln t}{N_x}}$
    \STATE \quad \quad\textbf{end for}
    \STATE \quad Select the configuration, ${x_t}^* = \arg\max_x UCB(x, t)$
    \STATE \quad Update the selection count $N_{{x^*,t}}$ = $N_{{x^*,t}}$ + 1
    \STATE \quad \textbf{end for}
    \STATE \textbf{return}  The optimal configuration ${x_{\text{opt}}} = \arg\max_x N_x$
  \end{algorithmic}
\end{algorithm}

Here, we present the details of \ouralg -- a lightweight online HPC application parameter selection algorithm specifically focusing on edge devices. 
The algorithm is developed based on the MAB framework and tailored to optimize scientific application configurations by balancing execution time and power consumption to facilitate user participation. It systematically explores a defined configuration space, $\chi$, that encompasses all possible combinations of input parameters for the application. The algorithm operates over a set number of iterations, $T$, and is calibrated using user-defined hyperparameters: weights for execution time ($\alpha$) and power consumption ($\beta$). These weights dictate the algorithm's balance of execution time minimization vs.\ power consumption reduction. We normalize the execution time ($\tau$) and power consumption ($\rho$) based on the MinMax normalization technique. 
% This normalization aligns the metrics uniformly, essential for their combined analysis. 
% The normalization process for each metric is defined as:
The normalized execution time $\tau_{\text{x}}$ is calculated as:
$\tau_{\text{x}} = \frac{\tau - \tau_{min}}{\tau_{max} - \tau_{min}}$, where $\tau_{min}$ and $\tau_{max}$ are minimum and maximum execution times, respectively. Similarly, the normalized power consumption $\rho_{\text{x}}$ is calculated as:
$\rho_{\text{x}} = \frac{\rho - \rho_{min}}{\rho_{max} - \rho_{min}}$, where $\rho_{min}$ and $\rho_{max}$ are minimum and maximum power consumption, respectively.
The weighted reward function, $f_{\text{reward}}(x)$, integrates the normalized execution time and power consumption values. The reward for selecting a configuration $x$ at iteration $t$, denoted as $R_{x,t}$, is determined as follows:
\begin{align}
f_{\text{reward}}(x) = \alpha \times \left( \frac{1}{\mu(\tau_{x})} \right) + \beta \times \left( \frac{1}{\mu(\rho_{x})} \right),
\label{eqn:reward}
\end{align}
where $R_x = f_{\text{reward}}(x)$ is the exploitation term, which is the weighted reward for configuration $x$. 
% $N_x$ is the count of times configuration $x$ has been selected up to iteration $t$. 
% Each configuration in the space $\chi$ is evaluated based on two key metrics: execution time ($\tau_x$) and power consumption ($\rho_x$). $\mu(\tau_x)$ and $\mu(\rho_x)$ represent the mean normalized execution time and mean normalized power consumption for a given configuration $x$, respectively. The algorithm operates over a set number of iterations, denoted as $T$. A calculated reward guides the selection, a function of the user-specified parameters $\alpha$ and $\beta$, as well as the normalized execution time and power consumption values.
Eq.~\ref{eqn:reward} ensures that the reward is inversely proportional to the normalized metrics of execution time and power consumption, thereby aligning with the user's optimization goals. 
The UCB in Alg.~\ref{alg:mab_algorithm} dynamically balances the exploration of new configurations against exploiting those already known to be effective. 
The performance of our algorithm is evaluated based on the total reward accrued over $T$ iterations. The expected total reward for a configuration $x$ is determined considering the randomness in execution time, power consumption, and the algorithm's selection strategy and is defined as follows:
\begin{align}
\mathbb E [R_x] = \mathbb E \left [\sum_{t=1}^{T} R_{x,t} \right].
\end{align}

% \begin{figure}[ht!]
% \centering
% \begin{tikzpicture}[node distance=0.6cm and 0.2cm, auto, scale=0.8, transform shape]
%     % Define styles
%     \tikzstyle{block} = [rectangle, draw, fill=blue!20, text width=4.5em, text centered, rounded corners, minimum height=3em]
%     \tikzstyle{line} = [draw, -latex']

%     % Place nodes
%     \node [block] (init) {Intialize Data};
%     \node [block, below=of init] (normalize) {Normalize Metrics};
%     \node [block, below=of normalize] (selectArm) {Select arm};
%     \node [block, below=of selectArm] (observeReward) {Observe Reward};
%     \node [block, below=of observeReward] (calcUCB) {Calculate UCB};
%     \node [block, below=of calcUCB] (select) {Get Max UCB};
%     \node [block, below=of select] (updateCount) {Update Count};
%     \node [block, below=of updateCount] (output) {Return Opt};

%     % Draw edges
%     \path [line] (init) -- (normalize);
%     \path [line] (normalize) -- (selectArm);
%     \path [line] (selectArm) -- (observeReward);
%     \path [line] (observeReward) -- (calcUCB);
%     \path [line] (calcUCB) -- (select);
%     \path [line] (select) -- (updateCount);
%     \path [line] (updateCount) -- (output);
%     % Add the additional edge from 'select' back to 'selectArm'
%     \path [line] (select.west) -| ++(-0.5cm,0) |- (selectArm.west);

% \end{tikzpicture}
% \caption{Multi-Armed Bandit Algorithm for Optimal Configuration Selection}
% \label{fig:mab_algorithm}
% \end{figure}

% For the UCB1 algorithm, which is a popular variant of UCB introduced by Auer et al. in 2002, 
The total regret \( R_n \) after \( n \) evaluations of a evaluations with \( K \) configurations is bounded by~\cite{slivkins2019introduction}:
% ~\cite{auer2002using}
\begin{align}
R_n \leq 8 \log(n) \sum_{i: \mu_i < \mu^*} \frac{1}{\Delta_i} + \left( 1 + \frac{\pi^2}{3} \right) \left( \sum_{i=1}^K \Delta_i \right),
\label{eq:bound}
\end{align}
where \( \mu^* \) denotes the highest expected reward (i.e., least execution time) among all configurations, \( \mu_i \) denotes the expected reward of the \( i \)-th configuration, and \( \Delta_i = \mu^* - \mu_i \) is the difference between the maximum expected reward and the reward of the \( i \)-th configuration.
The bound in Eq.~\ref{eq:bound} indicates that the regret grows logarithmically with the number of evaluations \( n \), which means that the average regret per play \( R_n / n \) tends to zero as \( n \) increases. This demonstrates the efficiency of the UCB-based approach in exploration-exploitation scenarios.

\subsection{\abrar{Integration with Existing Edge Computing Frameworks}}
\abrar{\ouralg integrates smoothly with existing edge computing frameworks due to its application-agnostic architecture and compatibility with protocols like CoAP (Constrained Application Protocol) \cite{rahman2016security}, enabling efficient communication and coordination between edge devices and HPC systems. However, challenges may arise from hardware differences, dynamic environments, and resource constraints on edge devices, particularly when tuning hardware-level parameters or maintaining real-time feedback. Addressing these requires careful protocol selection and configuration adjustments. As a modular algorithm, LASP can function independently or integrate with existing performance optimization components, as demonstrated in Section \ref{fig: comp with BLISS}, showing its effectiveness on devices with varying computational capabilities.}

\subsection{Challenges of the Proposed Approach} 

\abrar {

\textbf{Scalability Limitations:}
One limitation of \ouralg’s implementation is scalability. As the number of arms (configurations) increases, the UCB algorithm requires exploring a large number of options before it can intelligently determine the optimal configurations. This exploration becomes computationally intensive and inefficient, especially on resource-constrained edge devices.

\textbf{Network and Coordination issues:}
The presence of multiple volatile edge devices introduces additional challenges, particularly in terms of network issues. Low communication bandwidths between devices can hinder coordination and data transfer, impacting overall system efficiency.

\textbf{Scalability with Heterogeneous edge devices:}
One of the most complex challenges arises when scaling \ouralg to handle heterogeneous edge devices. These devices often have varying computational power, memory, and network connectivity, which can impact the effectiveness of a one-size-fits-all algorithm like UCB. Handling diverse device capabilities requires adaptive algorithms that can dynamically adjust resource consumption, depending on the device’s capabilities and environmental constraints. The varying performance characteristics across devices also increase the difficulty of ensuring that optimal configurations are found efficiently for each device. Future iterations of LASP will explore approaches like multi-level parallelism and resource-aware algorithm designs to better handle heterogeneous environments.}

\section{Evaluation}
\label{sec:evaluation}

Here, we first discuss details of \ouralg's execution, followed by performance evaluation against other configuration selection strategies. We then present how different user-level parameters affect \ouralg, and finally show how \ouralg can adapt to sensitivity changes.

\subsection{Experimental Setup}
We collected experiment data on the NVIDIA Jetson Nano device, a widely-used edge device in research~\cite{sk2022characterizing} and industry. The device's compact size, combined with its robust processing capabilities and power efficiency, makes the Jetson Nano a suitable choice for edge computing applications~\cite{lapegna2021clustering}. 

The Jetson Nano features a 128-core Maxwell GPU and a Quad-core ARM A57 CPU running at 1.43 GHz. It is optimized for efficient parallel processing and computation-intensive tasks. It runs on Ubuntu 20.04 OS and is equipped with 4 GB of 64-bit LPDDR4 RAM with a bandwidth of 25.6 GB/s. It uses a microSD card for storage. The device offers two power modes: MAXN and 5W. In Table \ref{tab:system_specs}, we provide a detailed description of each mode's specifications and operating characteristics.

\begin{table}[ht]
    \centering
    \caption{\abrar{System Specifications for MAXN and 5W Modes}}
    \begin{tabular}{|p{4cm}|p{1.5cm}|p{1.5cm}|}
        \hline
        \textbf{Parameter}               & \textbf{MAXN} & \textbf{5W} \\
        \hline
        \textbf{Power Budget (watts)}    & 10            & 5           \\
        \hline
        \textbf{Online CPU}              & 4             & 2           \\
        \hline
        \textbf{CPU Max Frequency (MHz)} & 1479          & 918         \\
        \hline
        \textbf{GPU TPC (MHz)}           & 921.6         & 640         \\
        \hline
    \end{tabular}
    \label{tab:system_specs}
\end{table}

This operational mode mimics the typical power constraints encountered in edge computing scenarios ~\cite{li2018end}. The high-fidelity data used in this study was collected on a system featuring an Intel® Core™ i7-14700 vPro® processor, with 20 cores and 28 threads, and a maximum turbo frequency of 5.3 GHz. The system had 64 GB of DDR5 memory  and ran on Ubuntu 24.04 LTS.

All the autotuning results and shown in the subsequent section are done on the Jetson Nano device to show the efficacy of our lightweight approach to autotuning. Furthermore, to mitigate potential performance interference, we ensured that no extraneous processes were running on the device, apart from the essential kernel processes and our target HPC applications.

\begin{table*}[t!]
\scriptsize
\centering
\caption{\abrar{HPC applications' configuration parameter ranges and their default values.}}
\label{table:default_application_configuration}
\vspace{-3mm}
\resizebox{1\linewidth}{!}{%
\begin{tabular}{|p{1.2cm}|p{5.8cm}|p{0.7cm}|p{2.5cm}|p{0.9cm}|}
\hline
\textbf{Application} & \textbf{Parameter Description} & \textbf{\abrar{Size}} & \textbf{Range} & \textbf{Default}\\
\hline
\multirow{3}{*}{kripke} 
& Layout: data layout and kernel implementation details & \multirow{3}{*}{\abrar{216}} & DGZ, DZG, GDZ, GZD, ZDG, ZGD & DGZ\\
& Gset: number of energy group sets &  & 1, 2, 3, 8, 16, 32 & 1\\
& Dset: number of direction sets &  & 8, 16, 32, 48, 64, 96 & 8\\
\hline
\multirow{2}{*}{lulesh} 
& r: number of regions to run for each domain & \multirow{2}{*}{\abrar{128}} & 1-15 & 11\\
& s: number of elements of cube mesh &  & 1-8 & 8\\
\hline
\multirow{3}{*}{clomp} 
& partsPerThread: \# of independent pieces of work per thread & \multirow{3}{*}{\abrar{125}} & 10, 20, 50, 70, 90 & 10\\
& zonesPerPart: number of zones &  & 100, 300, 500, 700, 900 & 100\\
& zoneSize: bytes in zone &  & 32, 128, 512, 1024, 2048 & 512\\
\hline
\multirow{9}{*}{hypre} 
& $P_x$, $P_y$: Processor grid size (x × y) & \multirow{9}{*}{\abrar{92160}} & 1 - 4 & 2\\
& strong\_threshold: AMG strength threshold &  & 0-1 & 0.25\\
& trunc\_factor: Truncation factor for interpolation &  & 1-10 & 2\\
& P\_max\_elmts: Max elements per row (AMG) &  & 1-4 & 1\\
& coarsen\_type: Algorithm for parallel coarsening &  & 1-3 & 1\\
& relax\_type: Defines which smoother to be used &  & 1-2 & 1\\
& smooth\_type: Number of smoothing levels &  & 0-1 & 0\\
& smooth\_num\_levels: Smoother level count &  & 1-4 & 3\\
& interp\_type: Parallel interpolation operator selection &  & 1-3 & 1\\
& agg\_num\_levels: Levels of aggressive coarsening applied &  & 1-10 & 2\\
\hline
\end{tabular}}
\end{table*}

\subsection{HPC Applications}
\abrar{Table~\ref{table:default_application_configuration} lists the HPC applications that we used to evaluate the effectiveness of our proposed techniques. To validate our results, we used applications with both small and larger parameter choices, excluding hardware-level parameters such as power and CPU capping. These applications cover a wide-ranging variety of science domains and have been used previously to capture the challenges in autotuning diverse HPC applications~\cite{marathe2017performance}.}

\textbf{\hypre}
\cite{falgout2002hypre} is a software library for scalable solutions of linear systems, leveraging parallel processing for high-performance computing. It includes the BLOPEX package for solving eigenvalue problems, making it a versatile tool for various scientific applications.

\textbf{\clomp}
~\cite{bronevetsky2009clomp} is a C-language benchmark that measures OpenMP overheads and performance impacts due to threading, simulating a typical scientific application inner loop workload under strong scaling conditions to assess the efficiency of various OpenMP scheduling algorithms.

\textbf{\lulesh}
~\cite{karlin2013lulesh} is a widely used proxy application that originated from the Shock Hydrodynamics Challenge Problem, designed to test the performance of high-performance computing systems and algorithms, and has since become a benchmark in DOE co-design efforts for exascale computing.

\textbf{\kripke}
~\cite{kunen2015kripke} is a scalable, 3D deterministic particle transport code that researches the effects of data layout, programming paradigms, and architectures on Sn transport implementation and performance, aiming to optimize solver performance and parallelism.

\subsection{Execution of \ouralg}

Here, we show how \ouralg finds optimal configuration using efficient parameter exploration, where we change the user's focus on controlling the optimization. We first show how \ouralg works when we have control over parameters in two dimensions for \lulesh. Next, we show the results for parameters in three dimensions for the application \kripke and \clomp. We also show the efficacy of \ouralg with multi-dimensional parameter application \hypre through our regret analysis and sampling efficiency to find the optimal configuration.

\begin{figure}[t!]
\subfigure[Power Focused]{\label{fig:lulesh_hm_t0.2}\includegraphics[width=0.23\textwidth]{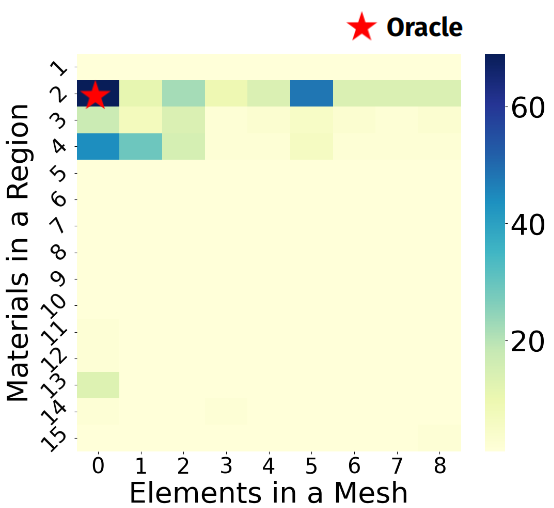}}\hspace{0.2cm}
\subfigure[Power Focused]{\label{fig:lulesh_hm_t0.4_time}\includegraphics[width=0.23\textwidth]{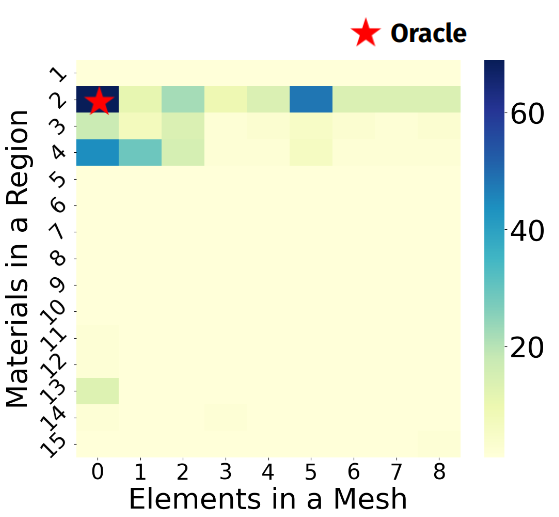}}
\subfigure[Time Focused]{\label{fig:lulesh_hm_t0.6}\includegraphics[width=0.23\textwidth]{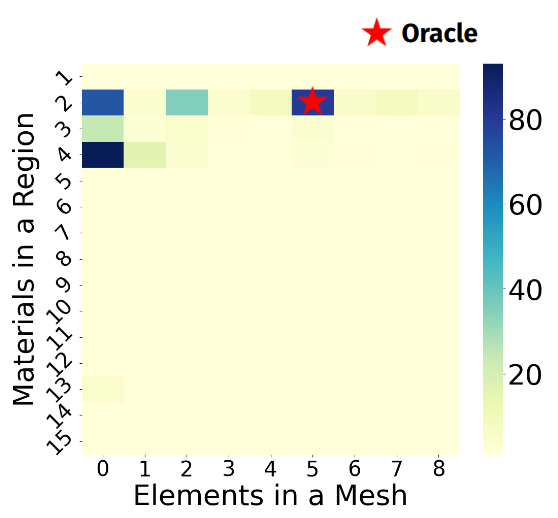}}\hspace{0.2cm}
\subfigure[Time Focused]{\label{fig:lulesh_hm_t0.8_time}\includegraphics[width=0.23\textwidth]{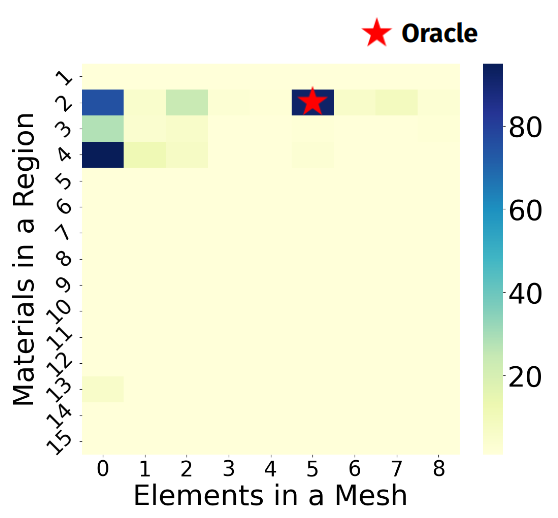}}
\vspace{-1mm}
\caption{(a) and (b) Exploration of the parameter space with Power as an objective metric for 1000 and 500 iterations, respectively. (c) and (d)  Exploration of the parameter space with execution time as an objective metric for 1000 and 500 iterations, respectively.}
\label{lulesh_heatmap}
\end{figure}

\textbf{Efficient Configuration Allocation:}
In {Fig.~\ref{lulesh_heatmap}}, we show how \ouralg achieves the optimal configuration. The figure presents a heatmap visualization of the configuration space for \lulesh, focusing on the application-level parameters ``Materials in Region'' and ``Elements in Mesh.''(The darker the cell, the more frequently LASP selected it as an optimal configuration.) The figure illustrates the frequency of the \ouralg's selection of specific configurations -- the darker regions indicating a higher selection frequency. We evaluated \ouralg over 1000 and 500 iterations, observing that in both scenarios, the algorithm effectively converges towards the optimal configuration. It is important to note, however, that the optimal configuration identified by \ouralg may not always be the most optimal, but close to optimal. This is due to \ouralg's stochastic nature, which navigates the configuration space based on the reward distribution of configurations. 
We adapted \ouralg to optimize both execution time and power consumption simultaneously. Fig.~\ref{lulesh_heatmap} shows that \ouralg effectively explores the configuration space, consistently identifying configurations that balance both objectives.
To test its efficiency, we ran \ouralg for 500 and 1000 iterations in two representative scenarios. Fig.~\ref{lulesh_heatmap} demonstrates that \ouralg converges to optimal configurations efficiently within 500 iterations when the parameter configuration dimensions are small (\lulesh, \kripke, \clomp). Whereas, running \ouralg for 1000 iterations helps it explore near-optimal configurations, which is beneficial for portability when deploying on traditional HPC clusters.

\begin{figure}[t!]
\subfigure[Time Focused]{\label{kripke_time0.8}\includegraphics[width=0.23\textwidth]{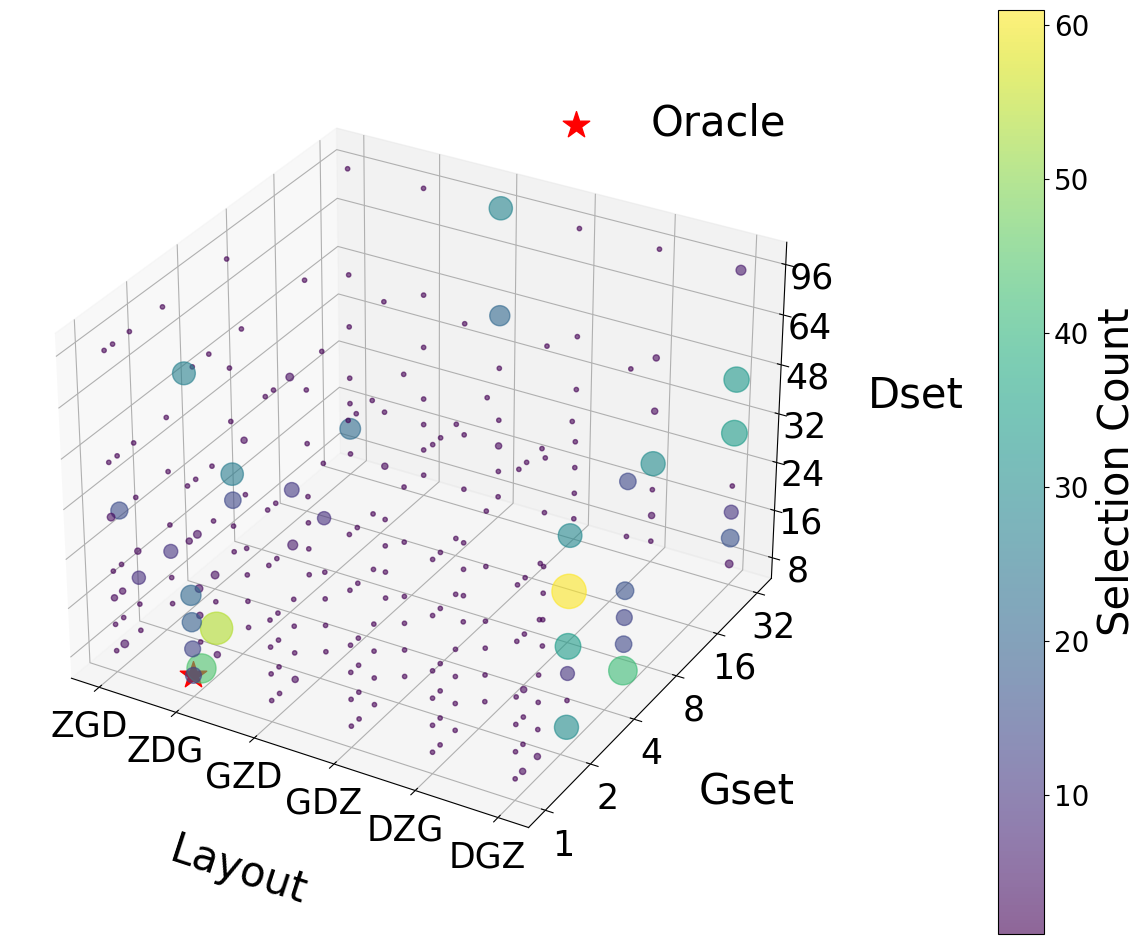}}\hspace{0.2cm}
\subfigure[Power Focused]{\label{kripke_power0.8}\includegraphics[width=0.23\textwidth]{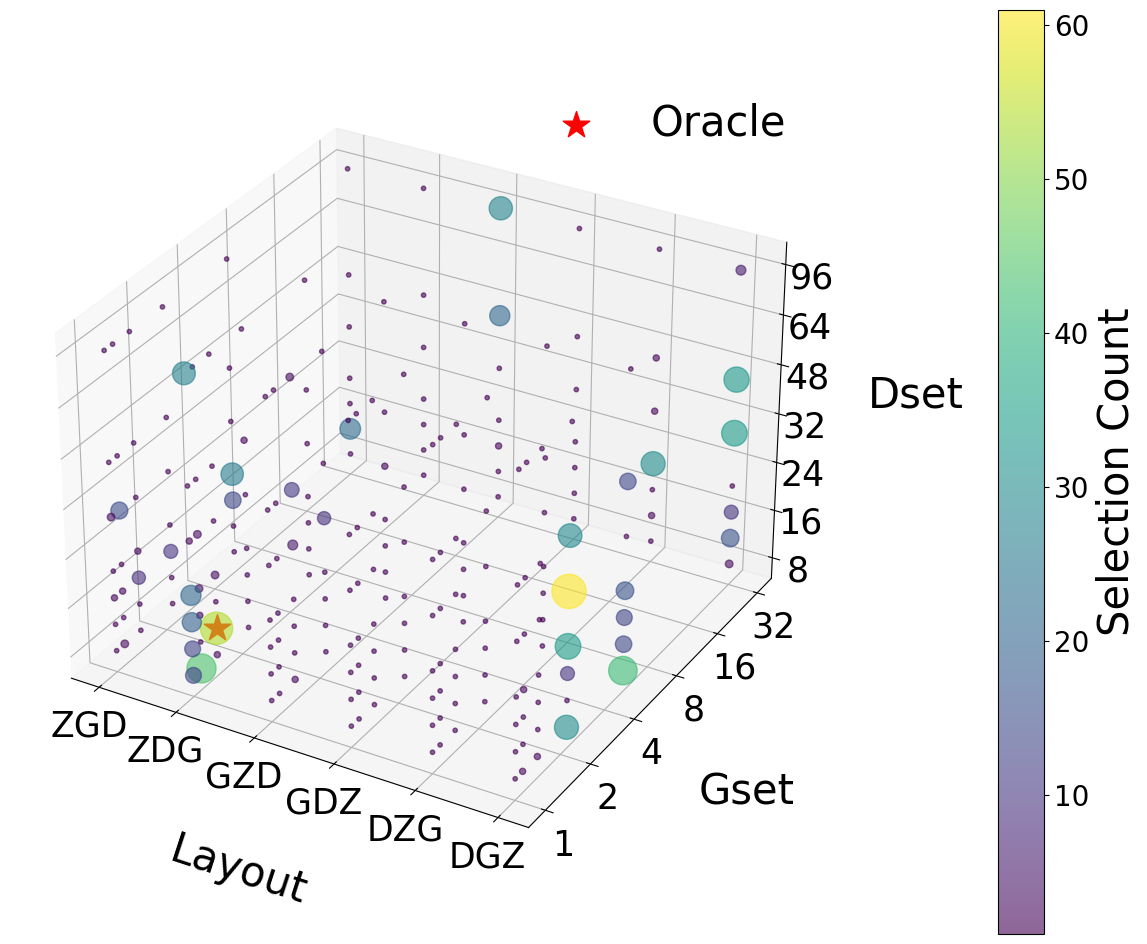}}
\subfigure[Time Focused]{\label{Clomp_time0.8}\includegraphics[width=0.23\textwidth]{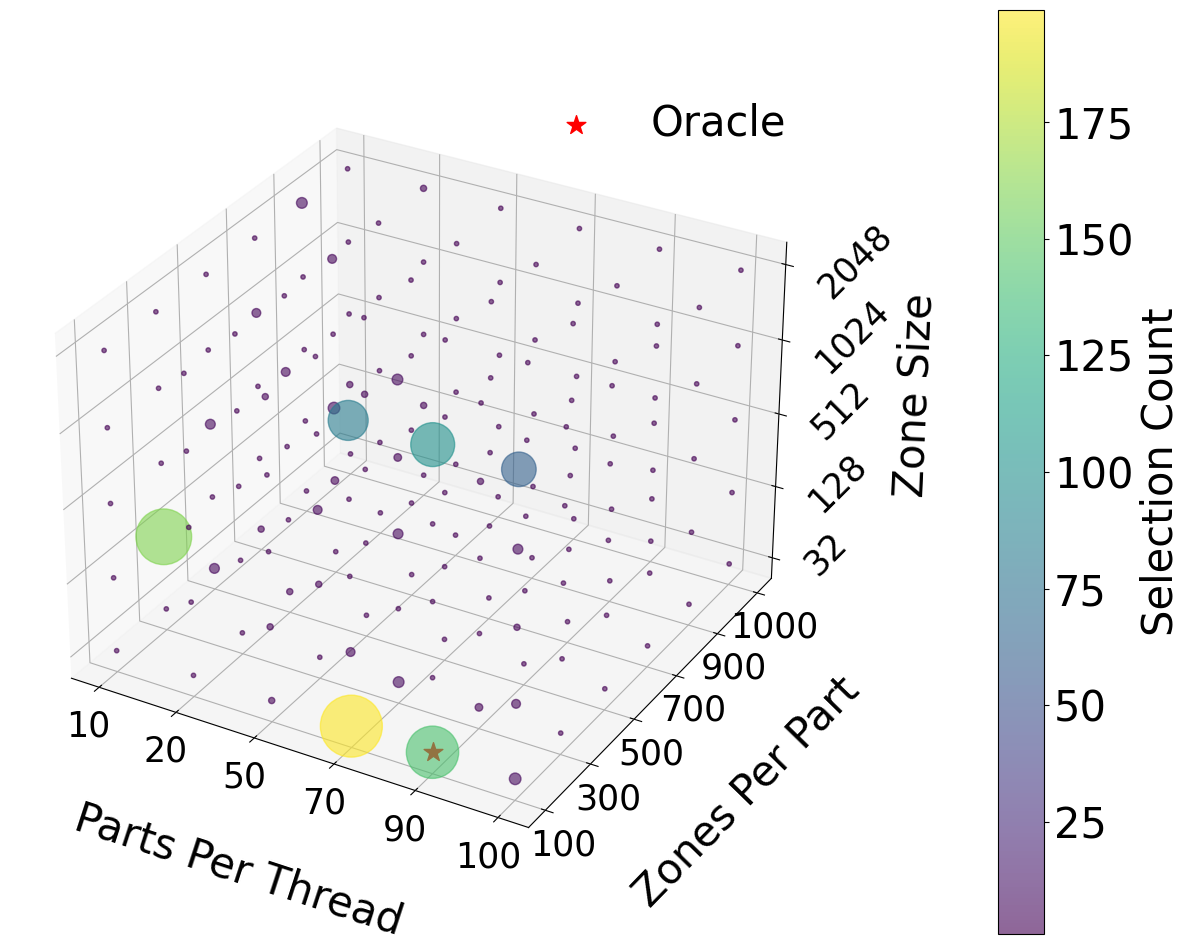}}\hspace{0.2cm}
\subfigure[Power Focused]{\label{Clomp_power0.8}\includegraphics[width=0.23\textwidth]{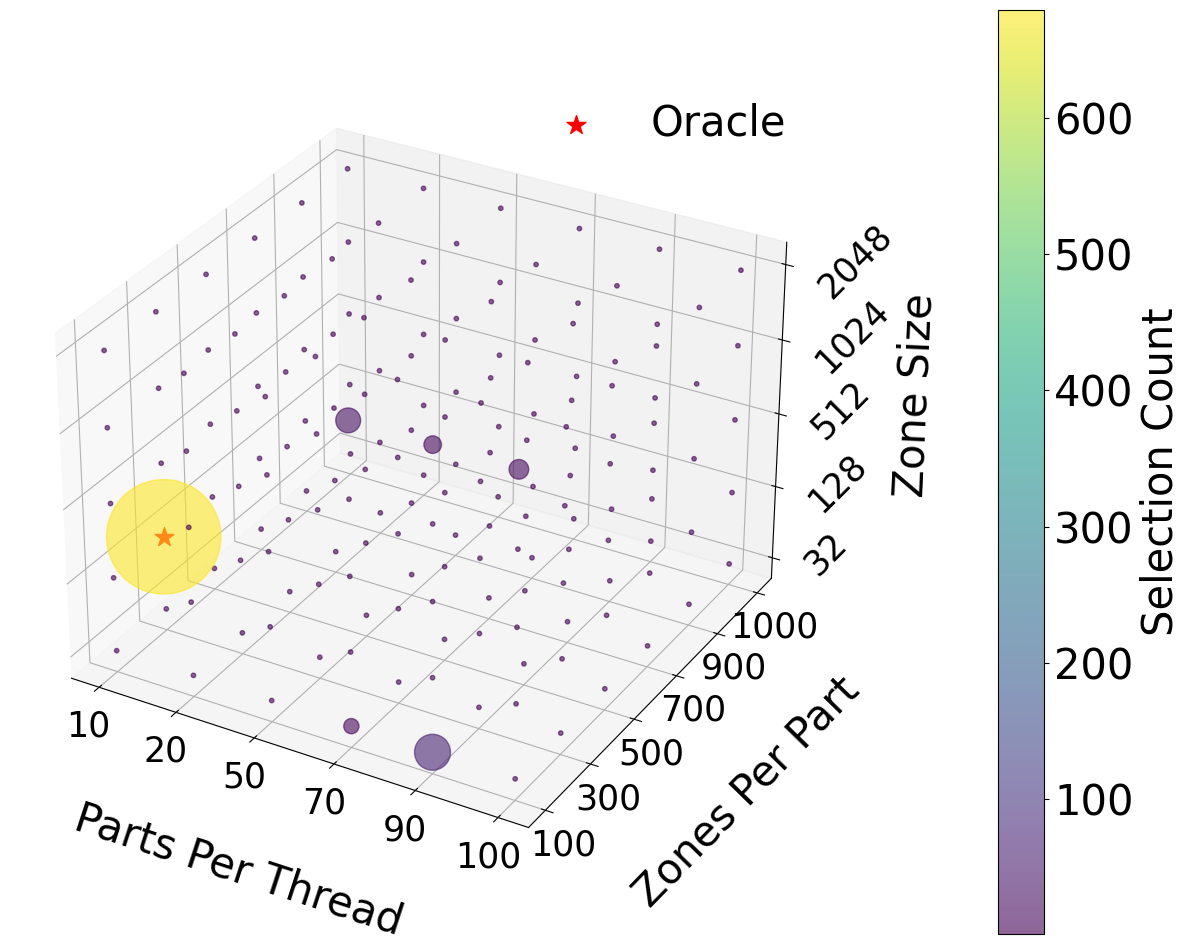}}
\vspace{-1mm}
\caption{Efficient exploration of the parameter space for Kripke (a \& b) and Clomp (c \& d).}
\label{kripke_clomp_heatmap}
\end{figure}

We performed a similar analysis for \kripke and \clomp, as shown in Fig.~\ref{kripke_clomp_heatmap}. This figure demonstrates the effectiveness and efficiency of \ouralg in high-dimensional parameter spaces. In Figs.~\ref{kripke_time0.8} and~\ref{kripke_power0.8}, we show how the optimal configuration is selected for \kripke in both time and power-focused experiments, respectively. Similarly, Figs.~\ref{Clomp_time0.8} and~\ref{Clomp_power0.8} illustrate the efficient convergence of the parameters for \clomp. In both cases, where execution time and power are used as objective metrics, \ouralg efficiently converges to the optimal configuration, as indicated by the oracle configuration.

\subsection{Performance Evaluation}
\label{sec:perf_evaluation}

%%%Time and power error graph

% (e.g., \kripke~\cite{kunen2015kripke}, \lulesh~\cite{karlin2013lulesh}, and \clomp~\cite{bronevetsky2009clomp}). 
The default values of parameters of these application have been shown in Table~\ref{table:default_application_configuration}.
% We take the default configurations for running the applications as baseline with which we compare the performance of our algorithm. 
% Table~\ref{table:default configuration} provides the range of the values for different applications and their default values. 
We calculate the performance gain under the best configuration \( PG_{\text{best}} \) as follows:
\begin{equation}
PG_{\text{best}} = \frac{f_{\text{default}} - f_{\text{best}}}{f_{\text{default}}} \cdot 100\%,
\label{eq:pg_best}
\end{equation}
where performance under default configuration is denoted as \( f_{\text{default}} \) and the performance under the best configuration is denoted as \( f_{\text{best}} \).

\begin{figure}[t!]
\centering
\subfigure[10W]{\label{fig:sampling_analysis_time}\includegraphics[width=0.48\columnwidth]{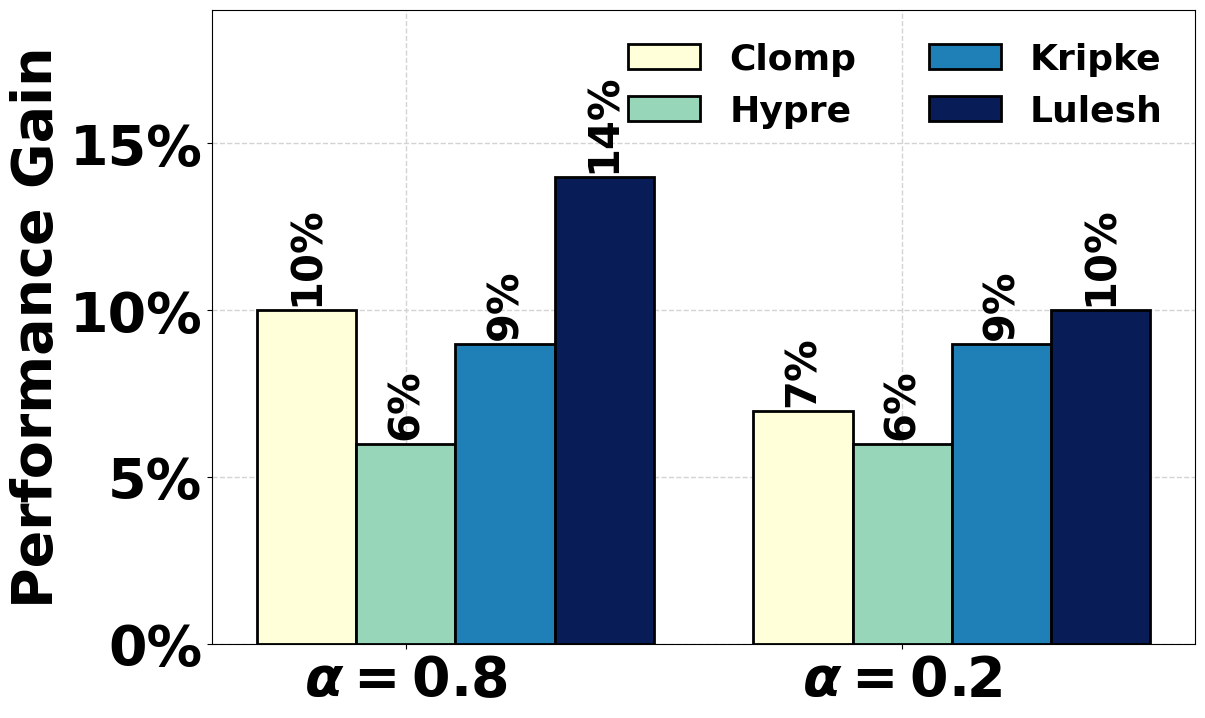}}
\vspace{-1mm}  
\subfigure[5W]{\label{fig:sampling_analysis_power}\includegraphics[width=0.48\columnwidth]{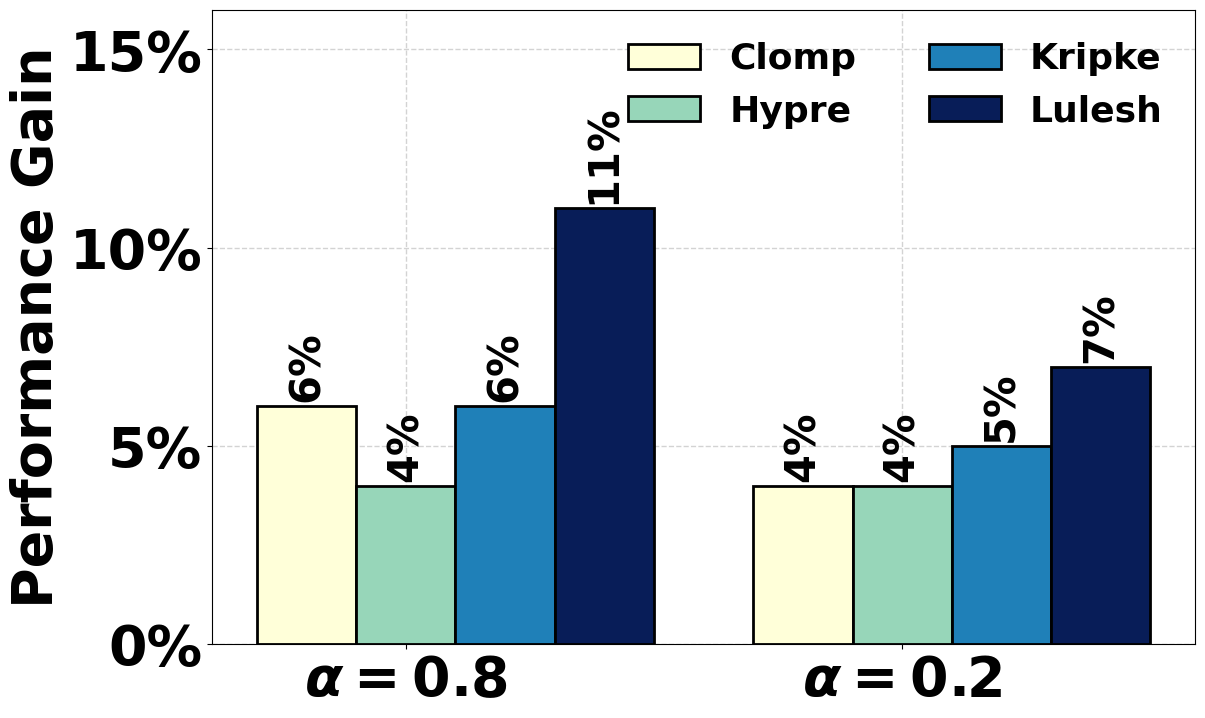}}
\vspace{-1mm}
\caption{Performance gain for different applications.}
\label{fig:perf_gain}
\end{figure}

In Fig.~\ref{fig:perf_gain}, we do this  performance gain analysis of the four applications by varying $\alpha$. 
At lower $\alpha$, \ouralg will work towards finding configurations with lower power consumption. When the user sets the power as the desired objective metric \ouralg achieves a 10\% performance gain for \clomp, 14\% for \lulesh,  9\% for \hypre and 6\%  for  \kripke. With increased $\alpha$, \ouralg will search the configuration space that yields lower execution time.

% \begin{figure}[t!]
%   \centering
%   \includegraphics[width=0.6\columnwidth]{figures/PG_graph_10W.png}
%   \vspace{-1mm}  
%   \caption{Performance gain for different applications.}
%   \label{fig:perf_gain}
% \end{figure}

%%%

%%%%
\ouralg achieves significant performance gains performance gain in execution time ($\alpha = 0.8$) and in power consumption with ($\alpha = 0.2$). As expected, \ouralg performs better in smaller parameter spaces compared to bigger ones, as shown in Fig.~\ref{fig:perf_gain}. This is because smaller parameter spaces allow for more efficient exploration and convergence to the optimal configuration.
\begin{figure}[t!]
\centering
\subfigure[Time Optimized]{\label{fig:sampling_analysis_time}\includegraphics[width=0.46\columnwidth]{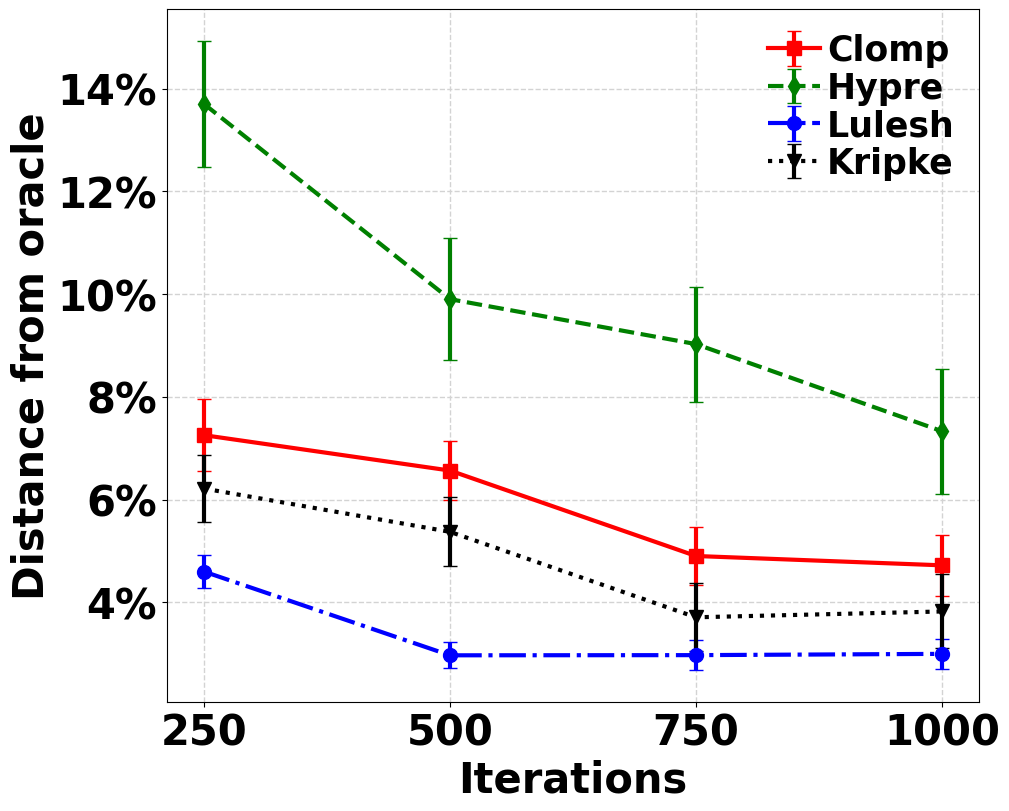}}
\subfigure[Power Optimized]{\label{fig:sampling_analysis_power}\includegraphics[width=0.46\columnwidth]{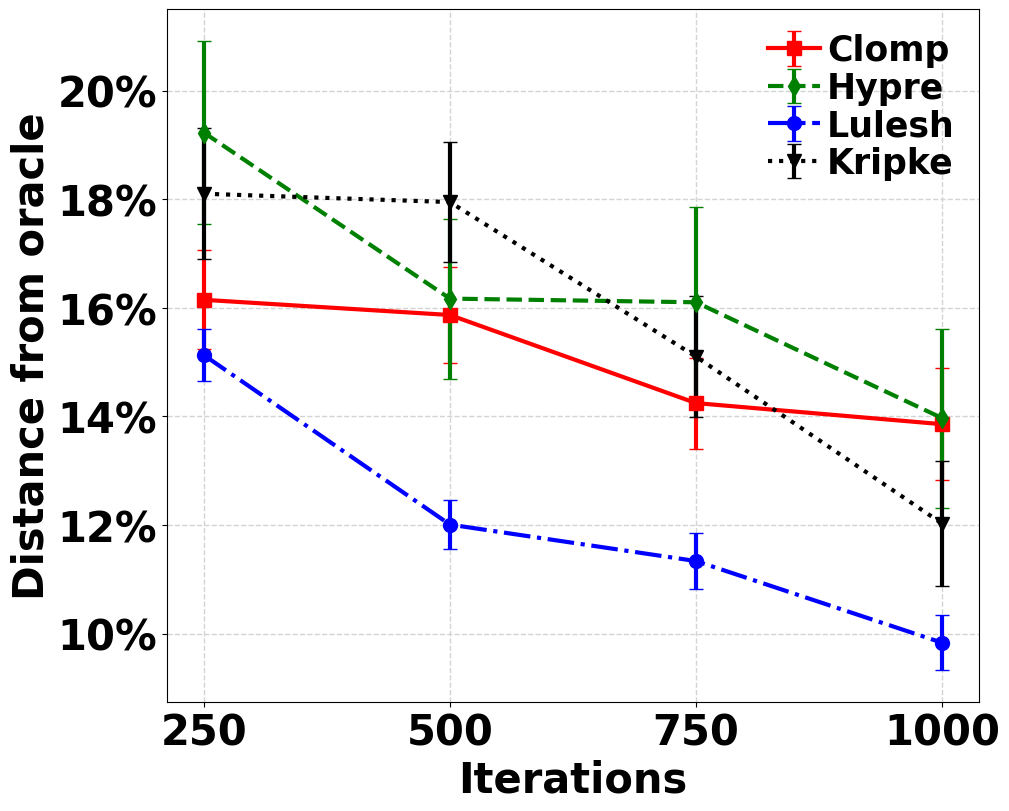}}
\vspace{-1mm}
\caption{\ouralg reaches closest to the Oracle with very few iterations.}
\label{sampling_analysis}
\end{figure}

However, \ouralg's fast convergence in finding the optimal configuration makes up for its performance in larger parameter spaces. We run \ouralg 100 times in order to see the mean distance from the oracle across different runs. The results are demonstrated in Fig.~\ref{sampling_analysis} which shows that \ouralg can reach within 12\% of the optimal configuration even in large parameter spaces, such as those of \hypre, when optimizing for execution time. When optimizing for power consumption, \ouralg's performance is less effective compared to when execution time as an objective metric. This is because power consumption is saturated by the edge device when running computationally intensive HPC applications, resulting in a less varied reward metric compared to execution time. As a result, \ouralg's ability to converge to the optimal configuration is impacted.

% \begin{figure}[t!]
% \centering
% \subfigure[CPU]{\label{fig:CPU}\includegraphics[width=0.46\columnwidth]{figures/CPU.png}}
% \subfigure[Memory]{\label{fig:Memory}\includegraphics[width=0.46\columnwidth]{figures/Memory.png}}
% \vspace{-1mm}
% \caption{\abrar{Resource Utilization for BLISS and LASP run on 10 W}}
% \label{fog:comparison with BLISS}
% \end{figure}

%%%

\begin{figure}[t!]
\subfigure[CPU (10W)]{\label{fig:10W_CPU}\includegraphics[width=0.23\textwidth]{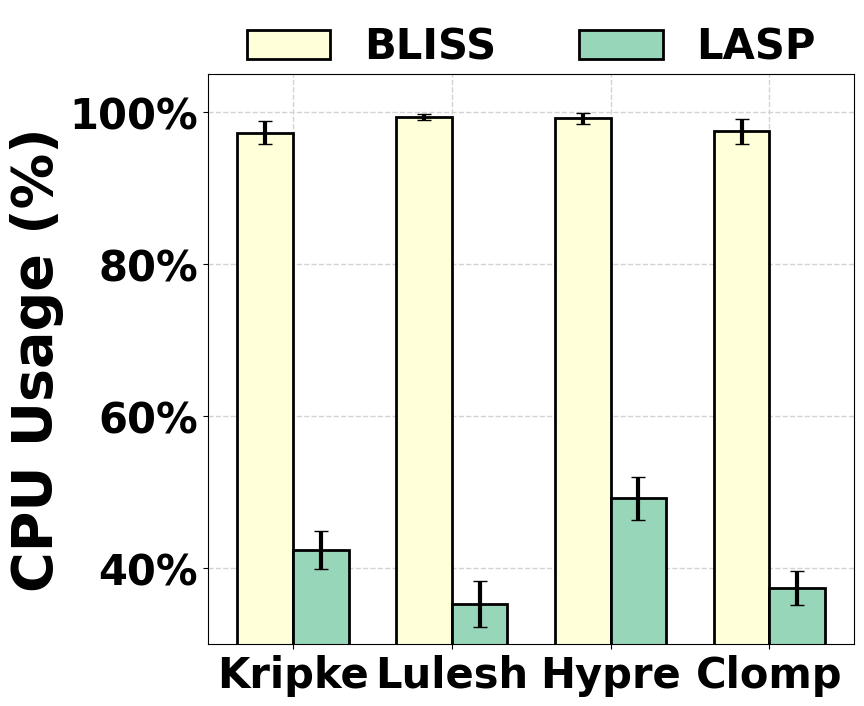}}\hspace{0.2cm}
\subfigure[Memory (10W)]{\label{fig:10W_mem}\includegraphics[width=0.23\textwidth]{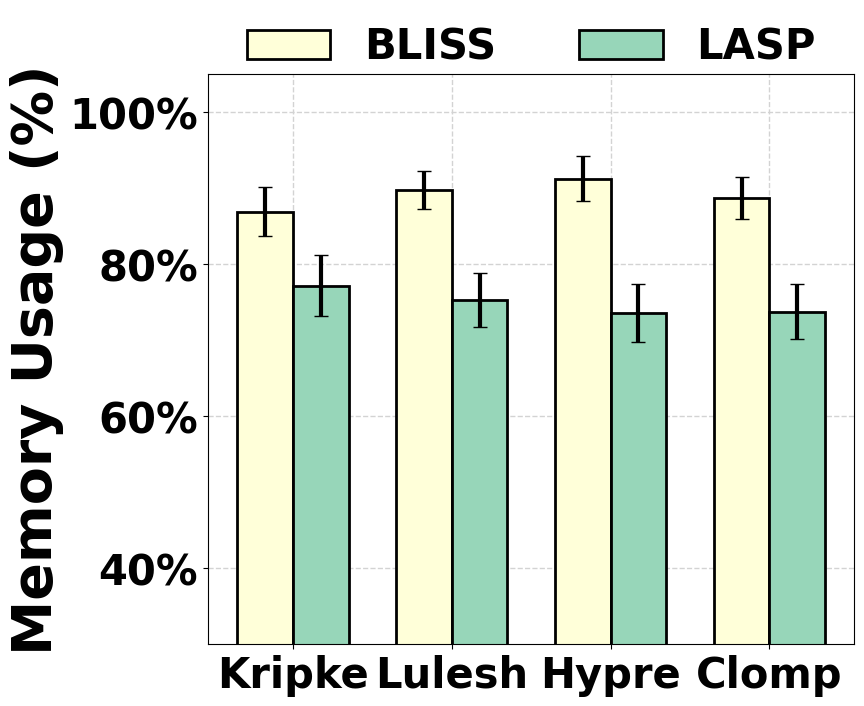}}
\subfigure[CPU (5W)]{\label{fig:5W_CPU}\includegraphics[width=0.23\textwidth]{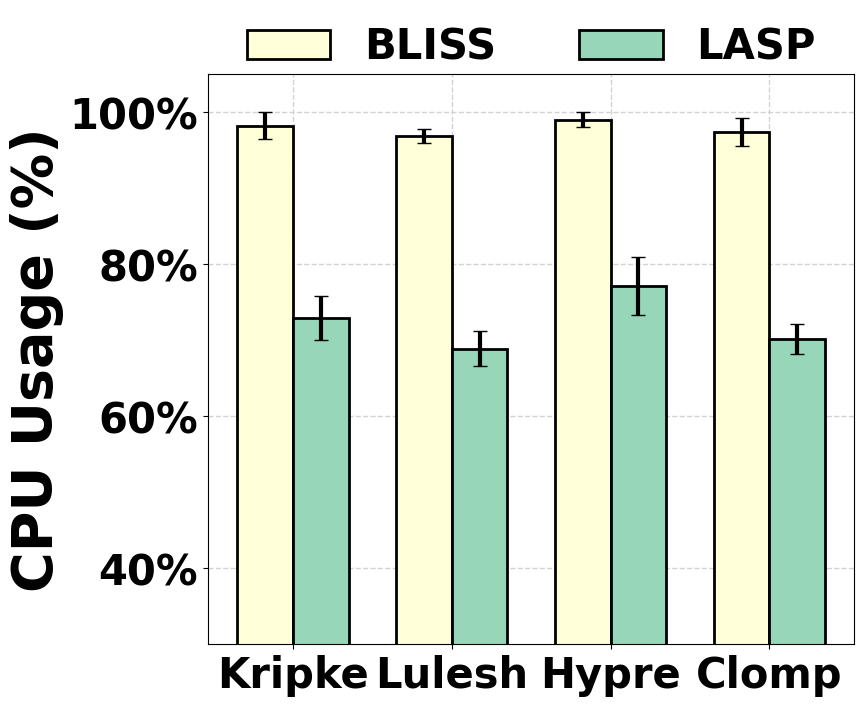}}\hspace{0.2cm}
\subfigure[Memory (5W)]{\label{fig:5W_mem}\includegraphics[width=0.23\textwidth]{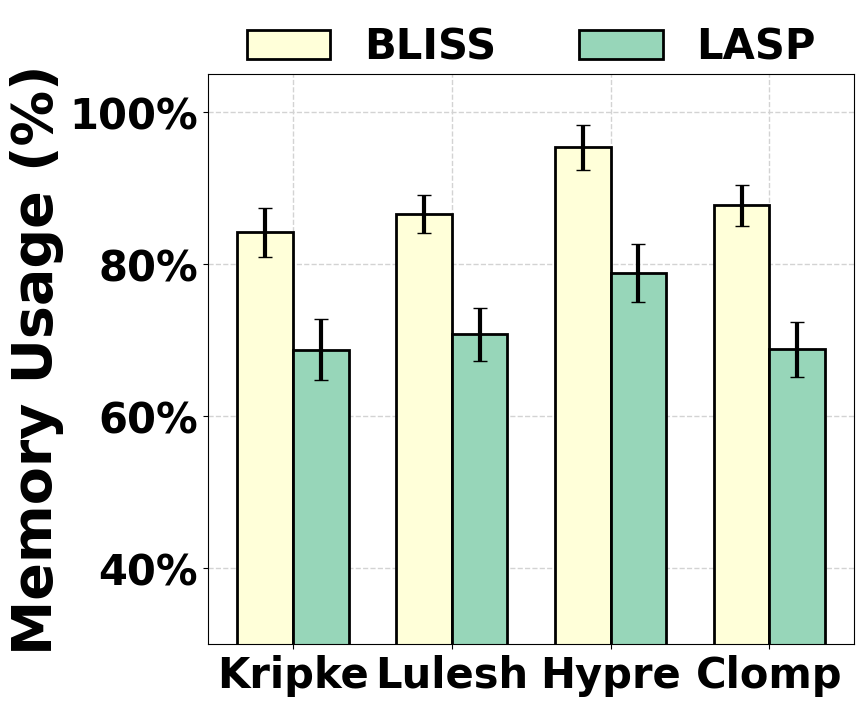}}
\vspace{-1mm}
\caption{{Resource Utilization of \ouralg compared to BLISS}}
\label{fig: comp with BLISS}
\end{figure}

\abrar{We compared our approach against BLISS \cite{roy2021bliss},a SOTA machine learning-based optimization method that leverages Bayesian Optimization (BO) to minimize tuning expenses. By creating a diverse pool of streamlined models, Bliss accelerates convergence and utilizes surrogate model predictions to streamline the evaluation of configurations, resulting in significant time savings. While we acknowledge our approach, did not do better in terms of efficiently finding the optimal parameters it is because we prioritized a lightweight approach for it to be applicable resource constrained edge devices. This is proved by our analysis of the CPU and memory footprint of using BLISS and LASP for autotuning on two modes (MAXN and 5W) to demonstrate the dynamic nature of our algorithm. A summary of our findings and a description of these two power modes are given in Fig \ref{fig: comp with BLISS}}.

%%%

\subsection{Regret Analysis}
\begin{figure}[t!]
\subfigure[\lulesh]{\label{fig:lulesh_regret}\includegraphics[width=0.23\textwidth]{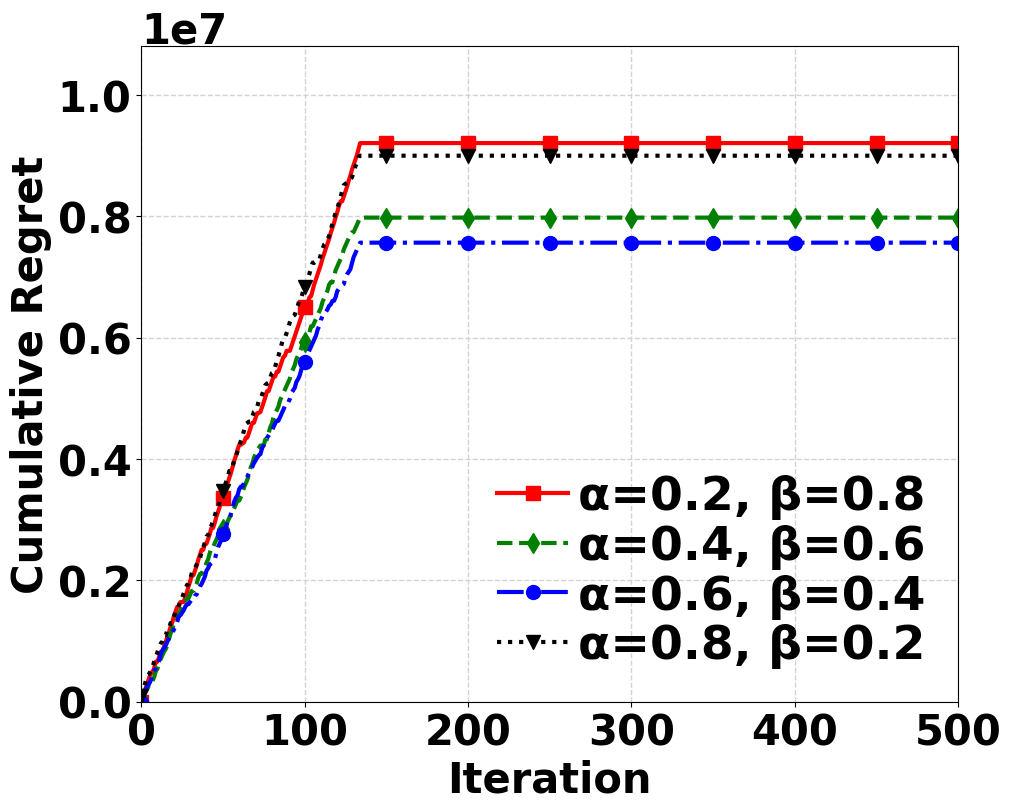}}\hspace{-3.2cm}
\subfigure[\kripke]{\label{fig:regret_plot}\includegraphics[width=0.23\textwidth]{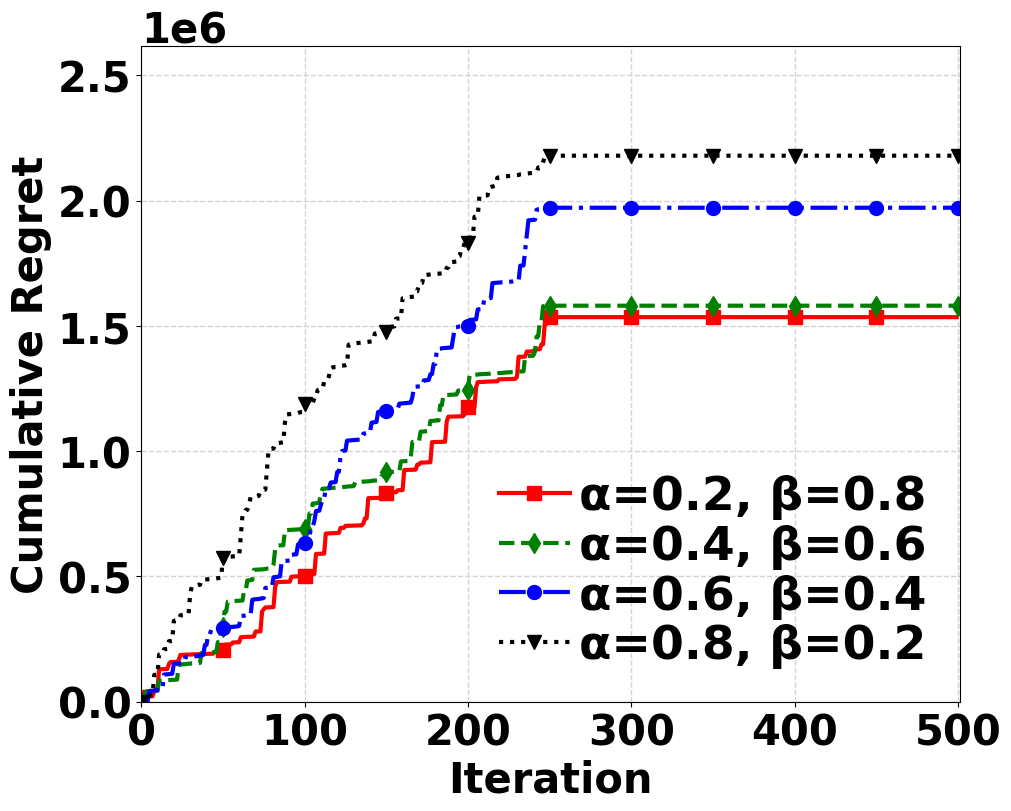}}
\subfigure[\clomp]{\label{fig:clomp_regret}\includegraphics[width=0.23\textwidth]{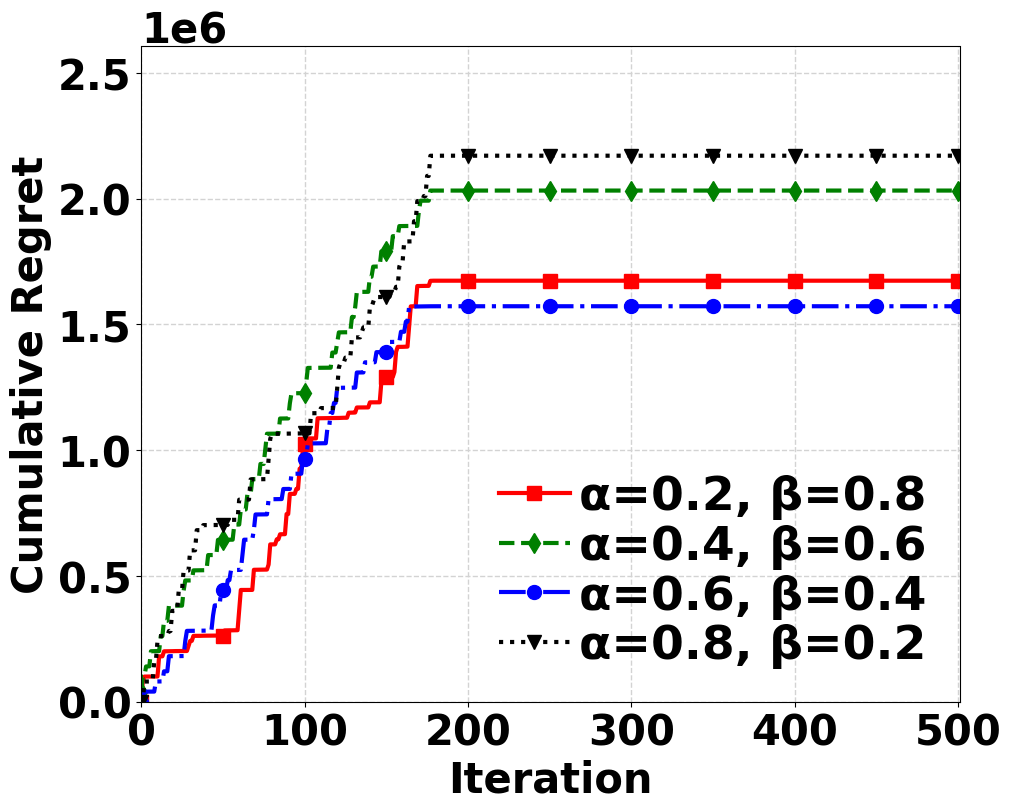}}\hspace{0.4cm}
\subfigure[\hypre]{\label{fig:hypre_regret}\includegraphics[width=0.23\textwidth]{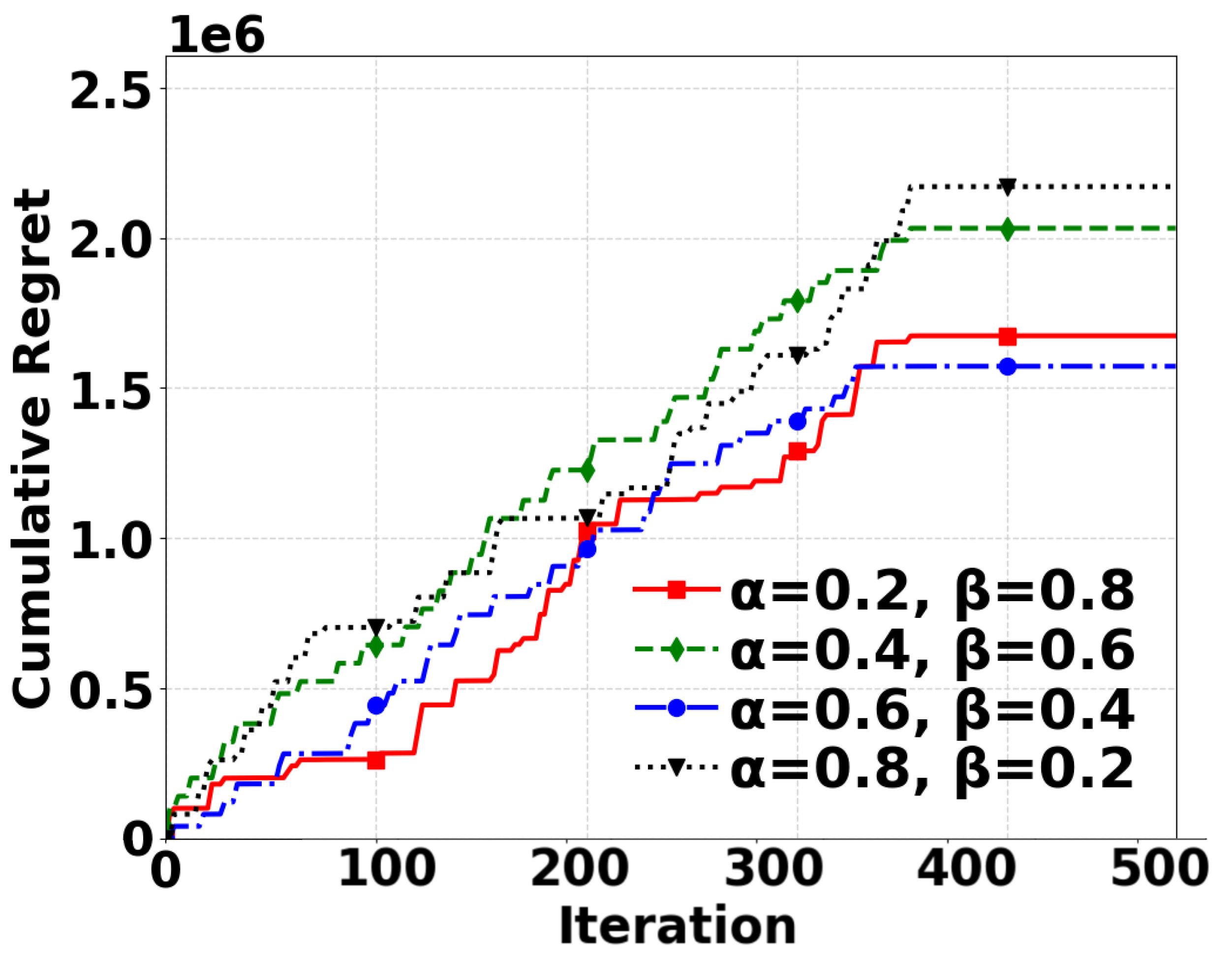}}
\vspace{-1mm}
\caption{Regret analysis for \lulesh, \kripke, \clomp and \hypre.}
\label{fig:regret_curves}
\end{figure}

We evaluate the efficiency of our proposed techniques by performing best-run(one time least regret run) regret analyses, as defined in Equation \ref{eq:regret_eqn}. The results, illustrated in Fig. \abrar{\ref{fig:regret_curves}}, showcase the convergence of \ouralg from an initial trial-and-error phase, characterized by suboptimal decision-making, to optimal configuration selection for four distinct applications.
By observing the accumulated regret at each iteration, we notice that the regret saturates after a certain number of iterations for all applications. The number of iterations required to reach minimal regret varies depending on the optimization metric. In the figures, we vary the value of 
$\alpha$
 from 0.8 (time-focused optimization) to 0.2 (power-focused optimization).
The plots reveal that \ouralg is more effective in finding configurations with shorter execution times. This is due to the variability of the collected data, which makes \ouralg better suited for optimizing execution times. As an online MAB-based technique, \ouralg navigates the search space based on environmental feedback.

\subsection{Sensitivity Study}

\begin{figure}[t!]
\centering
\subfigure[Time Optimized.]{\label{fig:time_optimized}\includegraphics[width=0.46\columnwidth]{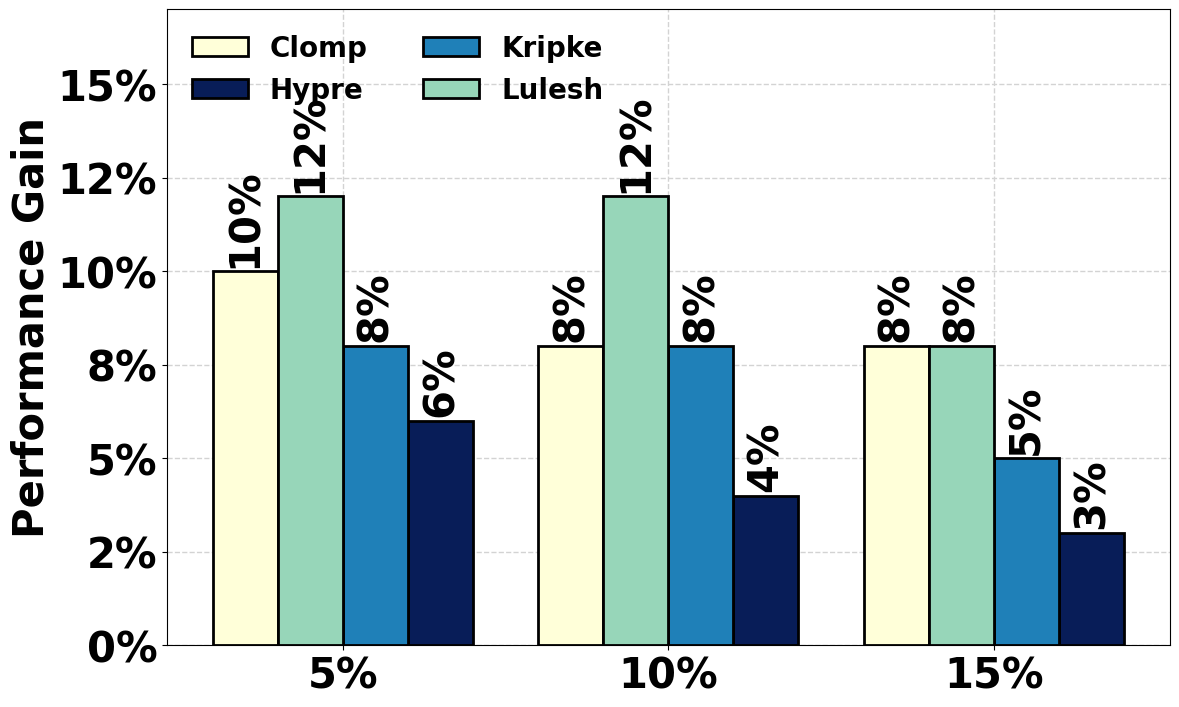}}
\subfigure[Power Optimized.]{\label{fig:power_optimized}\includegraphics[width=0.46\columnwidth]{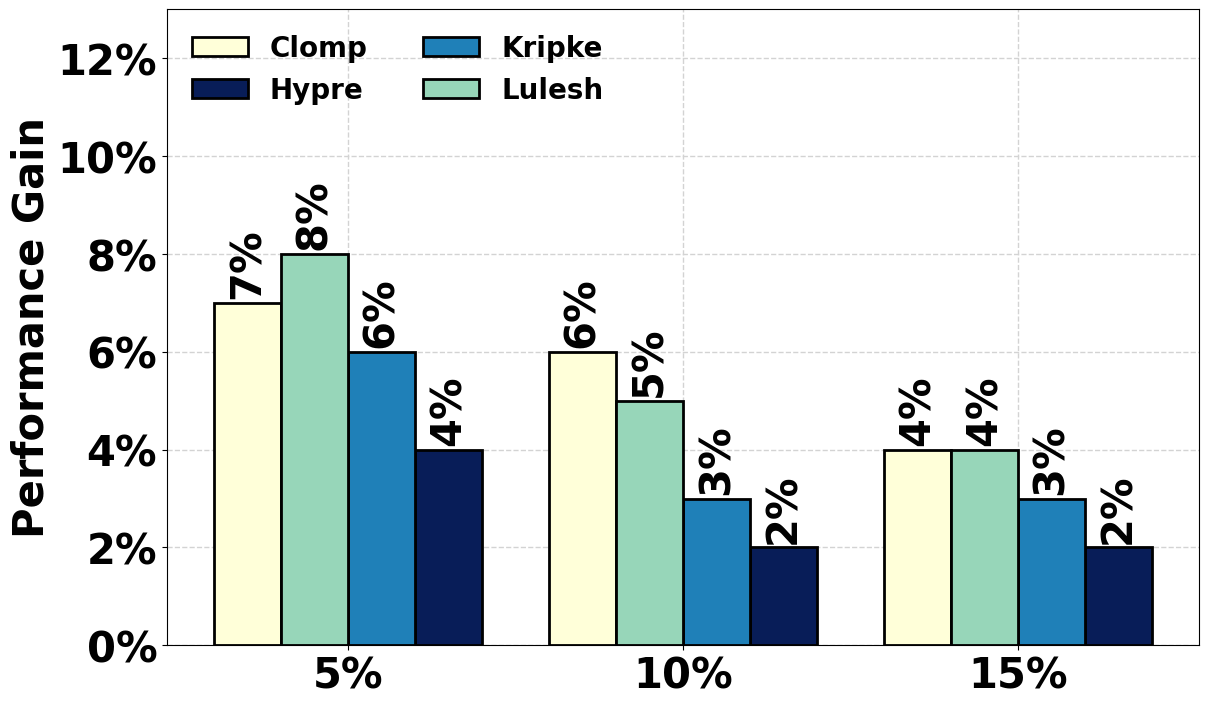}}
\vspace{-1mm}
\caption{Performance analysis with synthetic error in measurement data}
\label{error_analysis}
\end{figure}

\abrar{\textbf{Error in measurement data.} We introduce synthetic errors to the measured data to observe the dynamic nature of \ouralg. To simulate real-world imperfections, we add random noise to our collected data within a range of 5\%, 10\%, and 15\%. As can be seen in Fig.~\ref{error_analysis}, despite the erroneous feedback to \ouralg, we are still able to achieve considerable performance gains. This resilience can be attributed to the fact that MAB algorithms are inherently adaptive to change due to their design.

In this context, the random noise introduced in our experiments also serves as a proxy for network fluctuation anomalies, such as varying latencies or packet loss, which can lead to inconsistent measurements. Despite these additional challenges, \ouralg's ability to adapt to changing conditions allows it to mitigate the impact of such errors and continue to perform well even in the presence of network irregularities.}

\section{Concluding Remarks}
\label{sec:conclude}

In this paper, we introduce \ouralg, a novel and lightweight autotuning approach for dynamic configuration in resource-constrained edge systems. \ouralg stands out due to two key enhancements: firstly, it possesses the ability to learn and predict the configuration space in real-time, adapting swiftly to environmental changes. Secondly, it offers customization in optimizing both execution time and power consumption. To assess its effectiveness and efficiency, we conducted extensive experiments on four well-known HPC applications: \lulesh, \kripke, \clomp, and \hypre, each under varying settings. The results consistently demonstrated that 
\ouralg achieved a positive cumulative performance gain in dynamic workload scenarios. This capability is particularly beneficial for leveraging edge devices as proxies to perform the costly autotuning process. Our findings emphasize \ouralg's suitability for parameter tuning tasks, especially in environments where workloads frequently change.

\section{Acknowledgments}
This work is supported in part by the U.S. National Science Foundation under grants CNS-2300124, OAC-2411456, CCF-2324915, and ECCS-2152357. This work was performed under the auspices of the U.S. Department of Energy by Lawrence Livermore National Laboratory under Contract DE-AC52-07NA27344 (LLNL-CONF-855652).

% This work was performed under the auspices of the U.S. Department of Energy by Lawrence Livermore	National Laboratory	under Contract DE-AC52-07NA27344. (Add IM Release Number).

\bibliographystyle{ieeetr}
\bibliography{ref}

\end{document}